%% file: main.tex
\documentclass[aps,twocolumn,superscriptaddress,nofootinbib,noshowpacs,preprintnumbers,longbibliography,floatfix]{revtex4-1}
\usepackage[labelfont=bf,labelsep=quad,justification=centerlast]{caption}
\usepackage[utf8]{inputenc}
\usepackage{color}
\usepackage{graphicx}
\usepackage{amssymb,amsmath,amsthm,amsfonts}
\usepackage{mathtools,physics}
\usepackage{caption}
\usepackage{subcaption}
\usepackage{siunitx}
\usepackage{quantikz}
\usepackage[normalem]{ulem}
\usepackage{hyperref}
\usepackage{soul}
\usepackage{titlesec}
\titleformat*{\paragraph}{\itshape\bfseries}

\DeclareMathOperator*{\argmax}{arg\,max}

\begin{document}
\title{Modeling the Performance of Early Fault-Tolerant Quantum Algorithms}

\author{Qiyao Liang}
\affiliation{Zapata Computing, Inc.,
100 Federal Street, 20th Floor
Boston, MA 02110, USA}
\affiliation{Department of Electrical Engineering and Computer Science, Massachusetts Institute of Technology, Cambridge, MA 02139, USA}
\author{Yiqing Zhou}
\affiliation{Zapata Computing, Inc.,
100 Federal Street, 20th Floor
Boston, MA 02110, USA}
\affiliation{Laboratory of Atomic and Solid State Physics, Cornell University, Ithaca, NY 14853, USA}
\author{Archismita Dalal}
\affiliation{Zapata Computing Canada Inc., Toronto, ON, Canada, M5V 2Y1}
\author{Peter Johnson}
\affiliation{Zapata Computing, Inc.,
100 Federal Street, 20th Floor
Boston, MA 02110, USA}
\date{\today}

\begin{abstract}
Progress in fault-tolerant quantum computation (FTQC) has driven the pursuit of practical applications with \textit{early fault-tolerant quantum computers} (EFTQC). 
These devices, limited in their qubit counts and fault-tolerance capabilities, require algorithms that can accommodate some degrees of error, which are known as EFTQC algorithms. 
To predict the onset of early quantum advantage, a comprehensive methodology is needed to develop and analyze EFTQC algorithms, drawing insights from both the methodologies of noisy intermediate-scale quantum (NISQ) and traditional FTQC. To address this need, we propose such a methodology for modeling algorithm performance on EFTQC devices under varying degrees of error. As a case study, we apply our methodology to analyze the performance of Randomized Fourier Estimation (RFE)~\cite{kshirsagar2022proving}, an EFTQC algorithm for phase estimation. We investigate the runtime performance and the fault-tolerant overhead of RFE in comparison to the traditional quantum phase estimation algorithm. Our analysis reveals that RFE achieves significant savings in physical qubit counts while having a much higher runtime upper bound. We anticipate even greater physical qubit savings when considering more realistic assumptions about the performance of EFTQC devices. By providing insights into the performance trade-offs and resource requirements of EFTQC algorithms, our work contributes to the development of practical and efficient quantum computing solutions on the path to quantum advantage. 
\end{abstract}

\maketitle

\section{Introduction}
\input{v2/1_Intro}

\section{Randomized Fourier Estimation (RFE)}
\label{sec:rfe}
\input{v2/2_RFE}

\section{Framework}
\label{sec:roadmap}
\input{v2/3_Roadmap}

\section{Noise Modeling}
\label{sec:noise_model}
\input{v2/4_Noise}

\section{RFE Performance Analysis and Discussion}
\label{sec:analysis_discussion}
\input{v2/5_Performance}

\section{Conclusions and Outlook}
\label{sec:conclusion}
\input{v2/6_Conclusion}

\section*{Acknowledgements}
\label{sec:ack}
We thank Daniel Stilck Fran\c{c}a for helpful comments on the manuscript. We also thank Amara Katabarwa and Rutuja Kshirsagar for fruitful discussions that inspired this work.

\bibliographystyle{unsrt}
\bibliography{references}

\newpage
\onecolumngrid
\section*{Appendix}
\appendix
\input{Appendix}

\end{document}

%% file: v2/1_Intro.tex





Despite significant experimental and theoretical progress~\cite{arute2019quantum, kim2023evidence}, noisy intermediate-scale quantum (NISQ) devices have yet to exhibit the capacity to solve practical real-world problems with valuable outcomes.
A promising avenue towards achieving practical quantum advantage lies in the development of architectures that can support large-scale fault-tolerant quantum computations (FTQC)~\cite{nielsen2002quantum}. By incorporating robust fault-tolerance capabilities, we can suppress errors in our computations to an arbitrary extent. However, this comes at the cost of resources that far exceed the capabilities of present-day devices by several orders of magnitude. 
Projections by researchers indicate that millions to billions of physical qubits would be required to outperform classical computers in tasks such as factoring and ground state energy estimation~\cite{gidney2021factor, goings2022reliably, kim2022fault}.

There exists a substantial discrepancy between the 
capabilities of today's quantum devices and the projected resource requirements for practical large-scale fault-tolerant architectures. This discrepancy 
motivates
the question: \emph{how will the intermediate generation of devices, positioned between NISQ and FTQC, deliver practical advantage?} Such devices have recently been referred to as \emph{early fault-tolerant quantum computers} (EFTQC)~\cite{campbell2021early, tong2022designing}. 
Notably, EFTQC devices would possess a limited number of physical qubits, thus imposing constraints on the distance of the error-correcting codes they can support. These deviate from the conventional assumptions of fault-tolerant quantum computing, where such resources are presumed to be infinitely scalable.

A recent thrust in the field of quantum computing has been the development of EFTQC algorithms tailored to address the above limitations of EFTQC devices\mbox{\cite{campbell2021early, wang2021minimizing, katabarwa2021reducing,  lin2022heisenberg, zhang2022computing, tong2022designing, wan2021randomized, wang2022quantum, kshirsagar2022proving, giurgica2022low, wang2023faster, ding2023qcels, ni2023lowdepth, ding2023robust, Dalal_2023}}. So far, two key considerations are central to the pursuit of practical value using EFTQC algorithms. The first is developing quantum algorithms that reduce the number of qubits and operations per circuit, often at the expense of increased circuit runs and consequently extending the runtime\mbox{\cite{campbell2021early, lin2022heisenberg, zhang2022computing, wang2021minimizing, wang2023faster, ding2023qcels}}.
The second is designing quantum algorithms such that they are robust against gate and measurement errors\mbox{\cite{wang2021minimizing, katabarwa2021reducing, kshirsagar2022proving, Dalal_2023, ding2023robust}}.
These recent advancements showcase the potential applications of EFTQC devices, further motivating our
previously posed question. An essential next step is to develop methodologies for assessing the performance of these algorithms, enabling a deeper understanding of how intermediate devices between NISQ and FTQC can be leveraged to attain practical value.

In this work, we conduct an analytically case study, first to our best knowledge, that links logical circuit error models to the performance of a variant of the EFTQC algorithm mentioned above, the Robust Fourier Estimation (RFE) algorithm \mbox{\cite{kshirsagar2022proving}}. Specifically, we develop such a methodology to achieve the following:
\begin{itemize}
    \item A proof that quantifies the performance of the RFE algorithm that interpolates between using a single oracle call (suited for the high-noise NISQ setting) and using many oracle calls per circuit (suited for the low-noise FTQC setting) (see \S \ref{subsec:algperformance})
    \item A numerical demonstration of the suitability of this algorithm for EFTQC devices in that it can reduce by an order of magnitude the number of physical qubits required for large instances of phase estimation at the cost of an increase in runtime by several orders of magnitude (see \S \ref{subsec:ft_overhead})
\end{itemize}

To establish these results, we introduce a modularized methodology designed to assess the impact of generic circuit-level errors on a broad class of quantum algorithms. 
As a case study, we demonstrate the effectiveness of our methodology on the RFE algorithm, which shares a similar structure with other EFTQC-suited algorithms, such as those used for ground state energy estimation \mbox{\cite{lin2022heisenberg, wan2021randomized, wang2022quantum, ding2023qcels}} and property estimation \mbox{\cite{zhang2022computing}}.
The key feature common to these algorithms is the utilization of signals derived from Hadamard test outcomes, which makes our methodology applicable and adaptable to various cases within the EFTQC framework.


Our objective is to gain a comprehensive understanding of how errors impact the RFE algorithm. By doing so, we aim to leverage this understanding and extend it to algorithms designed for tasks beyond phase estimation. While previous studies have investigated the resilience of algorithms for quantum phase estimation \cite{kimmel2015robust}, we choose to analyze the RFE algorithm for two key reasons. Firstly, while sharing a similar structure to many of the above-mentioned EFTQC algorithms, the RFE algorithm is analytically tractable, making it an ideal candidate for our study. Secondly, its ability to smoothly interpolate between the high- to low-noise settings makes it well-suited for the EFTQC regime.  

The paper is organized as follows. We begin in \S \ref{sec:rfe} by introducing the RFE algorithm and our modifications to the algorithm for the purpose of this study. In \S \ref{sec:roadmap}, we outline the framework of our methodology, which we apply to analyze the runtime performance of the RFE algorithm. Specifically in \S \ref{sec:noise_model}, we develop a chain of noise models from physical errors to algorithmic errors, and study the impact of these errors on the performance of RFE in \S\ref{subsec:algperformance}. Based on these results, we provide a fault-tolerant resource estimation, comparing RFE to the traditional quantum phase estimation algorithm in \S\ref{subsec:ft_overhead}. Finally in \S\ref{sec:conclusion}, we conclude by highlighting our key results and provide an outlook of this work.

%% file: v2/2_RFE.tex
In this section, we introduce a variant of the RFE algorithm proposed in Ref.~\cite{kshirsagar2022proving}. 
The RFE algorithm is used to solve the task of phase estimation, in which the goal is to estimate the phase angle $\theta$ defined by $U|\psi\rangle = e^{i\theta}|\psi\rangle$, assuming the ability to prepare the eigenstate $\ket{\psi}$ and to implement controlled-unitaries $c$-$U$. The traditional approach to this problem is the quantum phase estimation (QPE) algorithm \cite{qpe}, which achieves the optimal performance asymptotically. However, the realization of the QPE algorithm requires multiple ancillary qubits and many high-fidelity quantum operations, both of which may be prohibitively-costly given hardware constraints in the early fault-tolerant regime. 

Given these constraints, alternative schemes for phase estimation have been proposed for near-term to early fault-tolerant devices \cite{wan2021randomized, lin2022heisenberg, dutkiewicz_heisenberg-limited_2022, obrien_quantum_2019}. In our study, we 
focus on the
RFE algorithm \cite{kshirsagar2022proving}, the performance of which can be analytically studied. Specifically, we will analyze a variant of the original RFE algorithm with slight modifications to simplify our analysis, as introduced later in this section. As the modifications made do not fundamentally change the mechanism of the algorithm, we will henceforth refer to the modified version of the algorithm as ``RFE" throughout the paper. 







\begin{figure}[!htb]
\centering
\begin{quantikz}
\lstick{$\ket{0}$} &  \gate[wires=1][0.1cm]{H}  &
\ctrl{1} & \gate[wires=1][0.1cm]{S(\phi)}\gategroup[wires=1,steps =3,style={dashed , rounded corners,fill=blue!20, inner xsep=2pt}, background]{Measurement in $\sigma_{\phi}$}&\gate[wires=1][0.1cm]{H} & \meter{} & 
\\
\lstick{$\ket{\psi}$} & \qw \qwbundle{}  & \gate{U^k} & \qw & \qw & &
\end{quantikz}
\caption{Diagram of the circuits used in the Randomized Fourier Estimation (RFE) algorithm. The parameter $k$ is uniformly randomly chosen among $\{0, \ldots, K-1\}$ for each iteration. A key feature of the algorithm is that the parameter $K$ controls the maximal circuit depth and is set to accommodate different degrees of error in the $c$-$U$ operation: high error implies small $K$ (low depth) and many repetitions, while low error warrants the use of large $K$ (high depth) and fewer repetitions. 
The boxed-up elements in blue can be collectively interpreted as a measurement with respect to the observable $\sigma_{\phi}=\cos(\phi)\sigma_x-\sin(\phi)\sigma_y$, where $\sigma_x$ and $\sigma_y$ are the conventional Pauli operators and $S(\phi)=\begin{bmatrix}
1 & 0\\
0 & \exp(i\phi)
\end{bmatrix}$.
}
\label{fig:hadamard_test}
\end{figure}
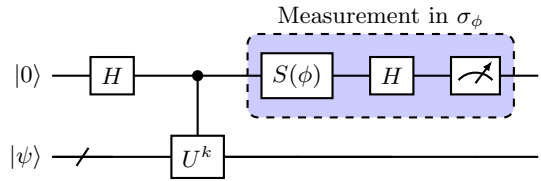 


The basic intuition behind this algorithm is summarized as follows. Each measurement outcome $z=\pm1$ is generated from a sampled parameterized Hadamard test circuit in Fig.~\ref{fig:hadamard_test} with $k\in\{0,\ldots,K-1\}$, where $K$ sets the maximal circuit depth, and $\phi\in[0,2\pi]$. 
\begin{figure}[!htb]
\includegraphics[width=0.48\textwidth]{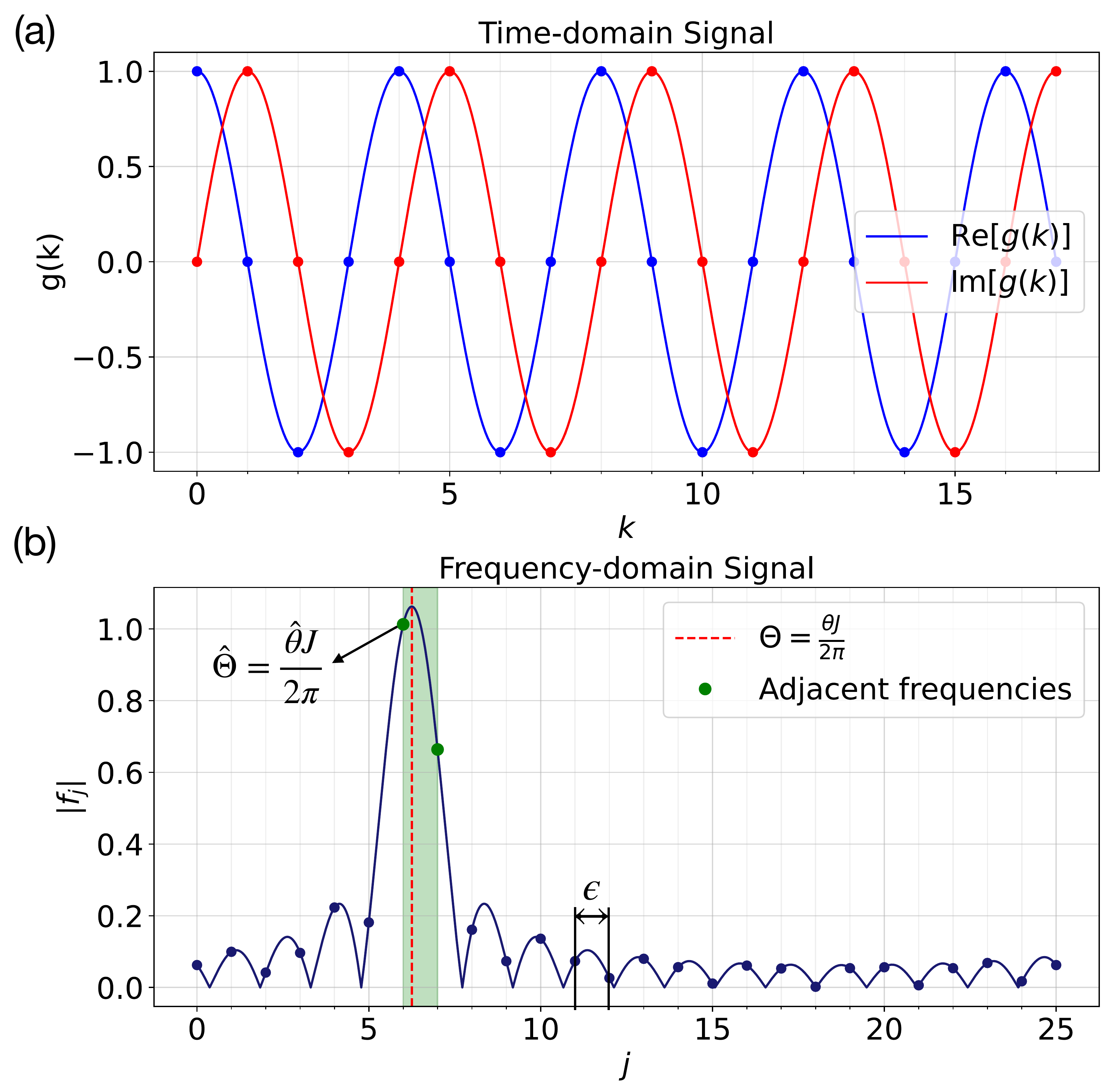}
\caption{\textbf{(a)} Real and imaginary components of the expected (noiseless) signal $\textup{Re}[g(k)]=\cos(k \theta )$  and $\textup{Im}[g(k)]=\sin(k \theta )$ as a function of the circuit depth $k$. \textbf{(b)} The magnitude of the expected Fourier transformed signal $|f_j|$. The green shaded region shows the acceptable range of values for $\hat{\theta}$, and the red dashed vertical line shows the true peak at index $\Theta=\frac{\theta J} {2\pi}$. The solid line represents the analytical functional forms where $k$ and $j$ are treated as continuous variables for easier visualization. The discrete values of $g(k)$ and $|f_j|$ are marked with dots. The two green dots in (b) correspond to $j=\lfloor J\theta/2\pi\rfloor$ and $\lceil J\theta/2\pi\rceil$, which we refer to as ``adjacent frequencies." }
\label{fig:noiseless_rfe}
\end{figure}
The expected outcome of the measurement $\langle\sigma_{\phi}\rangle=\cos(\phi)\langle \sigma_x\rangle-\sin(\phi)\langle\sigma_y\rangle = \cos(k\theta+\phi)$ is an oscillatory function of $k$ and $\theta$ and becomes the desired signal $\exp(ik\theta)$ when averaged over a uniform distribution of $\phi$ between 0 and $2\pi$ (i.e. $\frac{1}{2\pi}\int_0^{2\pi} 2\cos(k\theta+\phi)\exp(-i\phi)d\phi=\exp(ik\theta)$). 

To estimate the $\theta$ encoded in the frequency of the signal, we can construct from each outcome $z$ an unbiased estimate $\hat{f}_j$
of the expected discrete Fourier transform $f_j$ of signal $g(k)$ ($=\exp(ik\theta)$).
In the Fourier domain~$j\in\{0,\dots,J-1\}$, the estimate is expressed as
\begin{align}
\label{eq:estimator}
\hat{f}_j(k,\phi,z) = 2ze^{-i2\pi kj/J}e^{-i\phi}.
\end{align}
Given enough samples, we expect the magnitude of the averaged $\hat{f}_j$ to peak at a frequency close to~$\theta$. In the noiseless case, this peak will occur within the Fourier resolution $2\pi/J$ from the true $\theta$, where we later set the parameter $J$ such that the Fourier resolution matches the desired accuracy of the algorithm $\epsilon$. 

This peak frequency is then used as an estimate of $\theta$, denoted $\hat{\theta}$. Fig.~\ref{fig:noiseless_rfe}(a) shows the real and imaginary components of the signal $g(k)$ as a function of $k$.
The magnitude of the Fourier transformed signal $|f_j|$ is then plotted in Fig.~\ref{fig:noiseless_rfe}(b) as a function of $j$, where the peak occurs at an index near $\theta J/2\pi$ corresponding to the true frequency $\theta$.

The algorithm that we introduce here differs from that of Ref.~\cite{kshirsagar2022proving} in two regards. First, rather than taking samples corresponding to the real and imaginary parts of $g(k)$ separately by setting $\phi=0$ or $\pi/2$, $\phi$ is chosen uniformly randomly such that we can construct an unbiased estimator for $g(k)$ with a single shot. This simplifies the algorithm analysis without any changes in its performance. 

Second, the more substantial change is introduced to accommodate a more realistic noise model as developed in \S \ref{sec:noise_model}.
This noise model results in an exponential attenuation (as a function of $k$) of the outcome probabilities, converging towards a uniform two-outcome distribution at large $k$, similar to that of an unbiased coin toss. By appropriately setting the maximal value of $k$, i.e.~$K$, we can minimize the impact of this attenuation on our estimated $\hat{\theta}$ from the signal. 

In Ref.~\cite{kshirsagar2022proving}, the parameter $K$ was used to set both the maximum value of $k$ and the Fourier basis resolution $2\pi/K$.
This can be an issue in the case when the attenuation is strong and high accuracy is required: measurement outcomes for large $k$ are uninformative because they are drawn from a nearly uniform distribution. To address this issue, we allow these two values to differ; $K$ still labels the maximum value of $k$, while a new parameter $J$ is used to set the Fourier resolution. Then, the high accuracy and high noise case is accommodated by setting $J$ large and $K$ small. 
\begin{figure*}[!ht]
\includegraphics[width=\textwidth]{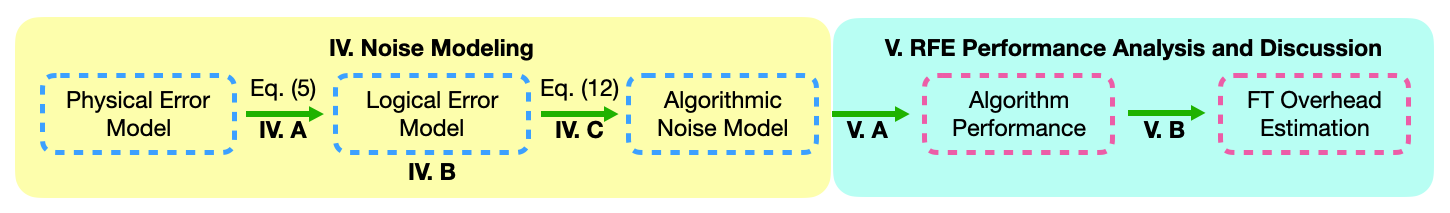}
\caption{Framework of our methodology.}
\label{fig:roadmap}
\end{figure*}

We now elaborate on why our algorithm works in the noiseless limit, which will provide intuition for its performance in the noisy case.
One can calculate the probability of measuring the outcome~$z$ in the Hadamard circuit of Fig.~\ref{fig:hadamard_test} as
\begin{align}
P(z|k,\phi;\theta)=\frac{1}{2}(1+z\cdot  \cos(k\theta+\phi)),
\label{eq:noiseless_lf}
\end{align}
which is an oscillatory function of $k$ with frequency $\theta$.
The two classically sampled variables are drawn uniformly: $P(k)=1/K$ and $P(\phi)=1/2\pi$, where the former distribution is discrete while the latter is continuous.
Using these distributions, the expected value of our constructed estimator~\eqref{eq:estimator} is calculated to be
\begin{widetext}
\begin{align}
\mathbb{E}[\hat{f}_j(k,\phi,z)]&=\sum_{k=0}^{K-1}\int d\phi \sum_{z =\pm 1}P(k) P(\phi)P(z|k,\phi;\theta)\hat{f}_j(k,\phi,z)=\frac{1}{K}\frac{1-a^K}{1-a}, \label{eq:expect_noiseless}
\end{align}
\end{widetext}
where $a:=e^{i\theta-i(2\pi j/J)}$. For $j\in \mathbb{R}^+$, the expectation~\eqref{eq:expect_noiseless} is maximized at $j = \theta J/2\pi$.
Whereas for $j\in \mathbb{Z}^+$, i.e.\ in a discrete Fourier transform setting, and assuming $K\le J$, the expectation achieves its maximum magnitude at $\lfloor\theta J/2\pi\rfloor$ or $\lceil\theta J/2\pi\rceil$ (or precisely at $\theta J/2\pi$ if $\theta J/2\pi\in \{0,\ldots, J-1\}$ ).
Consequently, by setting $J=2\pi/\epsilon$ we can then guarantee that the maximum of the expected discrete Fourier transform occurs at a frequency that is less than $\epsilon$ away from the true $\theta$. The maximal circuit depth
$K$ will be chosen according to the target accuracy $\epsilon$ and a parameter $\lambda$ (introduced in Eq.~\eqref{eq:algorithmic_noise_model}) that characterizes the error strength in the $c$-$U$ operation.
Qualitatively, $K$ is chosen to monotonically increase with $1/\epsilon$ and $1/\lambda$, which is described in further details in Section~\ref{subsec:algperformance}.

Our algorithm estimates the value of $\theta$ by using an average over multiple Fourier signal estimates~\eqref{eq:estimator}.
With $M$ estimates $\hat{f}_j^{(i)}$ generated independently, their average will concentrate about the expected value $f_j:=\mathbb{E}[\hat{f}_j(k,\phi,z)]$~\eqref{eq:expect_noiseless}, and a corresponding estimate of $\theta$ is given by
\begin{equation}
    \hat{\theta} = \frac{2\pi}{J}\argmax_j\left(\left|\frac{1}{M}\sum_i \hat{f}_j^{(i)}\right|\right).
    \label{eq:theta_estimate}
\end{equation}
This concentration will be addressed quantitatively in \S \ref{subsec:algperformance}.
The algorithm succeeds, i.e.\ yields an estimate such that $|\hat{\theta}-\theta|\leq \epsilon$, if 
one of the ``adjacent'' frequencies ($\lfloor \theta J/2\pi\rfloor$ or $\lceil \theta J/2\pi \rceil$) achieves the largest magnitude among all Fourier estimates $\hat{f}_j = \frac{1}{M}\sum_i\hat{f}_j^{(i)}$.
In expectation, one of these ``adjacent" frequencies (shown as the two green dots in Fig.~\ref{fig:noiseless_rfe}(b)) will achieve the largest magnitude, with a finite gap between it and the magnitudes of the non-adjacent frequencies (points that fall outside of the green shaded region). Hence, with sufficiently many samples $M$, the probability of failure of the algorithm can be made less than any finite failure probability $\delta$.

%% file: v2/3_Roadmap.tex
In this section we outline our proposed methodology for connecting the logical error model of an arbitrary quantum circuit to the success probability of a quantum algorithm, with the special case analysis of the RFE algorithm introduced in the last section as an example.

Our methodology, depicted in the flowchat of Fig.~\ref{fig:roadmap}, can be summarized as follows. We begin by modeling the effect of error on an algorithmic level from physical to logical-level error models proposed in \S \ref{sec:noise_model}. Specifically, in \S \ref{subsec:ft_model}, we establish a fault-tolerant overhead model that relates physical error rate $p_{\rm{phys}}$ to logical error rate $p_{\rm{logical}}$ for a surface code of distance $d$. We then propose a generic $N$-qubit logical Pauli error channel in \S \ref{subsec:logical_error_model} and statistically quantify its impact on an $N$-qubit quantum circuit of depth $D$. To achieve this, we assume that a random $N$-qubit Pauli error occurs after each layer of unitaries in our circuit such that the resulting state is a mixture of random states drawn from a unitary 2-design. By computing the expected value and variance of measurement outcomes based on this probability distribution, we gain insights into their statistical properties, which allow us to develop an algorithmic noise model in \S \ref{subsec:algorithmic_noise_model}.

In \S \ref{sec:analysis_discussion}, we investigate the performance of RFE under the algorithmic noise model proposed in \S \ref{subsec:algorithmic_noise_model}, namely, the exponential decay noise. In \S \ref{subsec:algperformance}, we give an upper bound on the algorithm runtime performance, and analyze its scaling with respect to the desired degree of accuracy $\epsilon$ in the presence of various strength of decay $\lambda$. Finally, in \S \ref{subsec:ft_overhead}, we provide a resource estimation of the RFE algorithm as compared to the standard QPE algorithm based on our original fault-tolerant overhead model proposed in \S \ref{subsec:ft_model}. 

%% file: v2/4_Noise.tex
\subsection{Fault-tolerant overhead model}
\label{subsec:ft_model}

We begin by establishing the connection between physical and logical error rates.
Currently, physical error rates range from $10^{-3}$ to $10^{-4}$, which are too large for reliable implementations of the RFE algorithm. 
To overcome this, we analyze the performance of our algorithm using lower logical error rates achievable through fault-tolerant computational protocols. 
Implementation errors at the circuit (logical) level arise from approximations in the operations (e.g., gate synthesis) and uncorrected errors in the fault-tolerant protocols.
The failure probabilities in both of these cases can be systematically reduced by paying a cost in the number of physical operations and, therefore, a cost in runtime.
In our analysis, we assume the operations to be non-approximate, focusing solely on the uncorrected errors in fault-tolerant protocols as the source of implementation error.

In order to quantify the reduction in error rates from the physical to logical level, we adopt the model proposed in Ref.~\cite{kim2022fault}.
This reduction in error rate comes at the cost of an increase in time and number of physical qubits; 
equivalently, these resources can be thought of as convertible into error rate reduction.
The quality of this conversion is governed by the ability of the particular architecture to maintain low physical gate error rates at scale.
A model for this conversion as a function of resource overhead~$d$ is expressed as
\begin{align}
 \label{eq:FT_model}   p_{\textup{logical}}=Ae^{-Bd},
\end{align}
where the parameters $A$ and $B$ depend on the physical error rates~\cite{kim2022fault}. 
The typical values for these parameters in the case of high (moderate) physical error rates are $A = 0.5 (0.4)$ and $B = 1.6 (1.1)$~\cite{kim2022fault}.
The overhead parameter $d$ can be thought of as the code distance in the context of a surface code~\cite{fowler2012surface}.
In this model, the physical qubit overhead is approximately $2d^2$.
In our subsequent analysis, we approximate the logical error model as a composition of single-qubit depolarizing errors~\eqref{eq:single_qubit_depolarizing}, where $p_\text{logical}$~\eqref{eq:FT_model} represents the depolarizing rate.
Based on these relationships, we can estimate the optimal performance of the RFE algorithm for given architecture parameters $A$ and $B$, as we will elaborate in \S\ref{subsec:algperformance}.

\subsection{Logical Gate Error Model}
\label{subsec:logical_error_model}

We consider a generic $N$-qubit Pauli error channel in order to study the impact of logical errors on the RFE algorithm.
A generic $N$-qubit Pauli error channel acting on an $N$-qubit density matrix~$\rho$ can be expressed by the following Kraus decomposition
\begin{equation}
    \Lambda(\rho)=\sum_{j=0}^{4^N-1} p_j A_j\rho A_j^{\dagger},
    \label{eq:n_qubit_pauli_kak}
\end{equation}
where $A_j\in\{I,X,Y,Z\}^{\otimes N}$ and $p_j$ is the probability of error $A_j$ occurring. We note that here we have assumed the most generic Pauli error channel for our logical errors to follow the convention of most error-correction literature, which can later be modified into different desirable error channels based on the set of $\{p_j\}$ of our choosing.

\begin{figure}[!htb]
\includegraphics[width=0.48\textwidth]{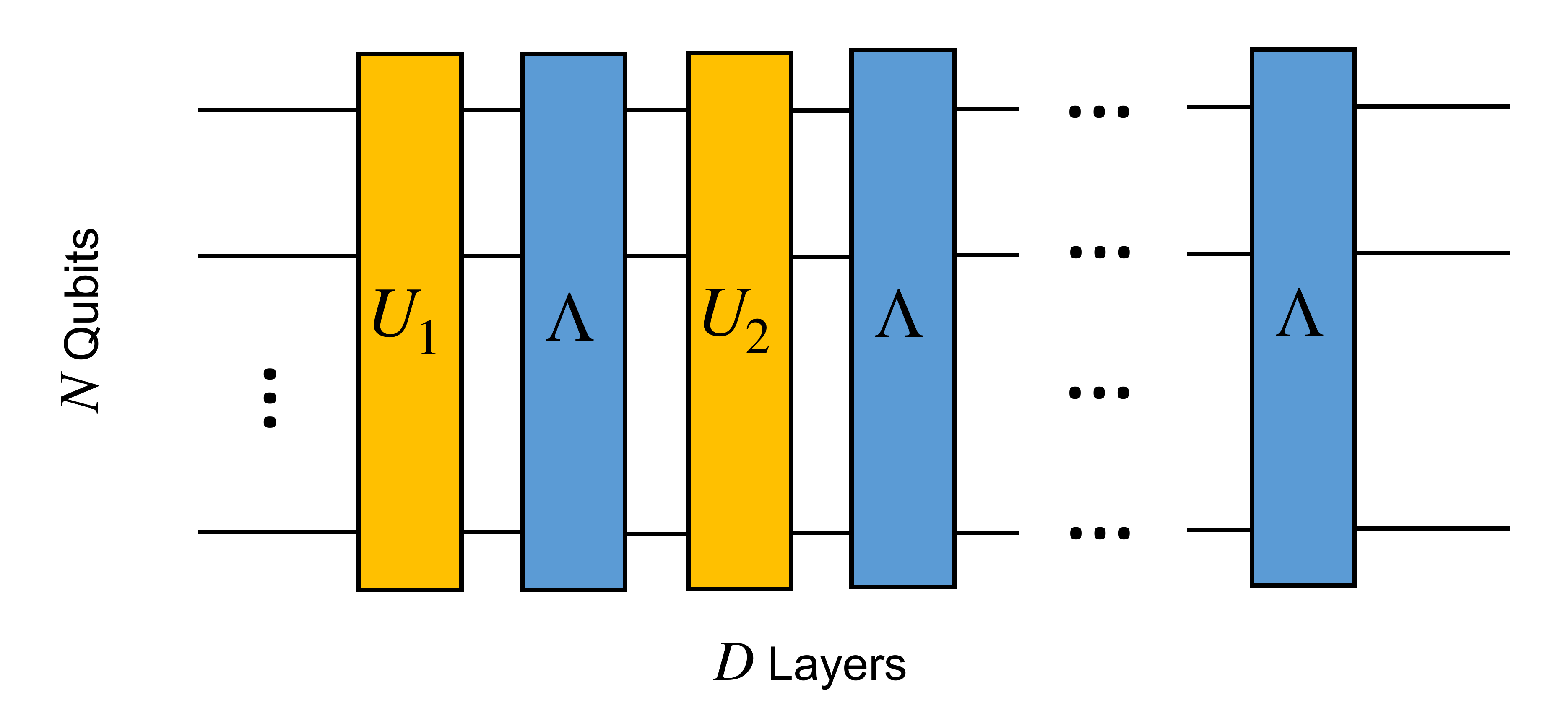}
\caption{Abstract circuit diagram of unitaries $U_1, \ldots, U_D$ interlaced with applications of a generic $N$-qubit Pauli error channel $\Lambda$ in an $N$-qubit $D$-layer random quantum circuit.}
\label{fig:circuit_diagram}
\end{figure}
Given a quantum circuit that implements the ideal unitary $U=U_D\cdots U_2U_1$ of depth $D$, we assume that the Pauli error $\Lambda$ interleaves the layers of unitary $\mathcal{U}_1$ through $\mathcal{U}_D$, where $\mathcal{U}_i(\rho):=U_i\rho U_i^{\dagger}$ is the superoperator describing the action of the unitary $U_i$ on the density operator $\rho$. The overall noisy circuit is illustrated in Fig.~\ref{fig:circuit_diagram}. The outcome state $\rho_f$ upon applying this noisy circuit to the initial state $\rho_i$ is given by
\begin{widetext}
\begin{align}
\rho_f =
\Lambda \circ\mathcal{U}_D\circ\cdots \circ \Lambda \circ\mathcal{U}_1\left(\rho_i\right) = \sum_{j_D=0}^{4^N-1}\cdots \sum_{j_1=0}^{4^N-1}p_{j_D}\cdots p_{j_1} (A_{j_D}U_D\cdots A_{j_1}U_1)\rho_i(U_1^{\dagger}A_{j_1}^{\dagger}\cdots U_D^{\dagger}A_{j_D}^{\dagger}). \label{eq:kraus_product}
\end{align}
\end{widetext}

After obtaining the noisy output quantum state~\eqref{eq:kraus_product}, we now focus on the various statistical properties of the final measurement with respect to an N-qubit Pauli observable.
Here we introduce the notion of a ``trajectory state," defined as $\ket{\psi_{\mathbf{j}}}\equiv\ket{\psi_{j_1\ldots j_D}}:= A_{j_D}U_D\ldots A_{j_1}U_1\ket{0^N}$, where $\mathbf{j}$ is the index tuple that includes all of $j_1,\ldots, j_D$. Each of these trajectory states $\ket{\psi_{\mathbf{j}}}$ corresponds to one combination of errors occurring on $\rho_i$ among the $4^{ND}$ possibilities with probability $p_{\mathbf{j}}=p_{j_1}\ldots p_{j_D}$. Expressing the expectation of an observable~$P$ with respect to the state $\rho_f$ in Eq.~\eqref{eq:kraus_product} in terms of these trajectory states, we get
\begin{align}
    \langle P\rangle &= \Tr\left[\Lambda \circ\mathcal{U}_D\circ\cdots \circ \Lambda \circ\mathcal{U}_1\left(\rho_i\right)P\right]\\
    &= p_{\mathbf{0}}\Tr\left[\ket{\psi_{\mathbf{0}}}\bra{\psi_{\mathbf{0}}}P\right]+\sum_{\mathbf{j}=1}^{4^{ND}-1}p_{\mathbf{j}}\Tr\left[|\psi_{\mathbf{j}}\rangle\langle\psi_{\mathbf{j}}|P\right],
    \label{eq:quantum_exp}
\end{align}
where $\ket{\psi_{\mathbf{0}}} = U_D\ldots U_1\ket{0^N}$ is the ideal state upon which no error has occurred. 
To provide a tractable analysis of $\langle P\rangle$, we propose the 
\emph{unitary 2-design model}. 
Under this approximation, we replace each of the noisy trajectory states $|\psi_{\mathbf{j}}\rangle$ with a randomly sampled state from a spherical 2-design \cite{gross2007evenly} (i.e. any distribution over pure states whose first and second moments match those of the Haar distribution of unitaries over $N$-qubits applied to a reference state).
This assumption establishes the statistics of $\langle P\rangle$, allowing us to compute its mean and variance as
\begin{align}
    \mathbb{E}[\langle P\rangle]&= p_{\mathbf{0}}\langle\psi_{\mathbf{0}}|P|\psi_{\mathbf{0}}\rangle \text{ and} \label{eq:quantum_exp} \\
    \text{Var}[\langle P\rangle]&= \frac{1}{2^{N}+1}\sum_{{\mathbf{j}}=1}^{4^{ND}-1}p_{\mathbf{j}}^2
    \label{eq:quantum_var},
\end{align} 
respectively.
By our assumption, each of these noisy trajectories shares the same mean. Consequently, as an increasing number of trajectories are averaged over to form the second term in Eq.~\eqref{eq:quantum_exp}, their collective average will converge towards this mean.
A detailed calculation of the above quantities is provided in Appendix \S\ref{app:stats}.
We would also like to point out that the unitary 2-design model proposed here to establish Eq.~{\eqref{eq:quantum_exp}} and {\eqref{eq:quantum_var}} assumes that the noisy trajectory states to be drawn uniformly randomly from all possible states, which is a strong assumption for errors happening in a worst-case scenario. An improved version of the model would likely benefit from system-specific knowledge of noises in the early fault-tolerant devices and their corresponding distributions of noisy trajectory states, which warrants further investigation. 

Lastly, we briefly mention the potential impact of state preparation errors on our analysis. Small amounts of error in the initial state will not significantly affect the results of our paper. However, larger amounts of error can have a more pronounced effect. For example, if the initial state is not perfectly prepared in the ground state, then the peak in {Fig.~\ref{fig:noiseless_rfe}} will be suppressed in height and additional peaks may appear, each with a height proportional to the overlap between the imperfect initial state $|\tilde{\psi}\rangle$ and other eigenstates $|\psi\rangle$ of $U$, i.e. $|\langle \tilde{\psi}| \psi\rangle|^2<1$. To avoid over-complicating the analysis, we choose to omit the consideration of state preparation errors in our noise modeling. We further note that a recent study {\cite{ni2023lowdepth}} investigates the performance of a similar-spirited EFTQC algorithm for phase estimation as RFE under the effect of imperfect state preparation. Combining our work with the analysis of Ref.{\cite{ni2023lowdepth}} would provide additional insights into the interplay between different error sources in the early fault-tolerant regime.

\subsection{Logical Gate Error Model to Algorithmic Noise Model}
\label{subsec:algorithmic_noise_model}

Based on the statistics of $\langle P\rangle$ from Equations \eqref{eq:quantum_exp} and \eqref{eq:quantum_var}, we propose an algorithmic noise model that captures the effect of noise as a combination of exponential decay and random fluctuations on our algorithm.
The exponential decay term stems from the $p_{\mathbf{0}}$ term in the mean of $\langle P\rangle$, whereas the random fluctuation is related to the variance of $\langle P\rangle$.
For the RFE algorithm, the impact of error is to alter the probability~\eqref{eq:noiseless_lf} of the measurement outcome~$z$ as 
\begin{align}
    \textup{Pr}(z|k,\phi;\theta) = \frac{1}{2}(1+ze^{-\lambda k} \cos(k\theta -\phi)+\eta_{k,\phi}),
    \label{eq:algorithmic_noise_model}
\end{align}
where we introduce $\eta_{k,\phi}$ to represent a noise bias, i.e.\ random fluctuations, and $\lambda$ to parameterize an exponential attenuation of outcome probability with respect to circuit depth $k$. Letting $\eta_{k,\phi}$ and $\lambda$ vary arbitrarily, this ``algorithmic" noise model is completely general and the algorithm would clearly not succeed in all settings.
The key to establishing a reasonable algorithm performance is to limit, at least statistically, the magnitude of $\eta_{k,\phi}$ under different strengths of $\lambda$, as we elaborate later in this subsection.

We now analyze how RFE works in a noisy setting under our proposed algorithmic noise model \eqref{eq:algorithmic_noise_model}.
Given, for example, a single-qubit depolarizing error channel
\begin{equation}
    \mathcal{D}(\rho) = (1-r)\rho+\frac{r}{3}(X\rho X+Y\rho Y+Z\rho Z),   \label{eq:single_qubit_depolarizing}
\end{equation}
the $N$-qubit composite error channel is given by
\begin{equation}
    \Lambda(\rho) = \mathcal{D}^{\otimes N}(\rho),
\label{eq:n_qubit_concatenated_single_q_depolarizing_noise_channel}
\end{equation}
a special case of the $N$-qubit Pauli error channel from Eq.~\eqref{eq:n_qubit_pauli_kak}.
Upon applying the operation~c-$U$, where $U$ is assumed to have $D$ layers, $k$ repetitions in the Hadamard test circuit, the probability of the signal remaining noiseless at the end is
\begin{align}
    p_{\rm{total}} = p_{\mathbf{0}}^k = (1-r)^{NDk},
\end{align}
which decreases as a function of c-$U$ depth number $k$. 
From Eq.~\eqref{eq:quantum_exp}, we learn that the expected value of the quantum expectation value $\langle P\rangle$ decays with the total probability $p_{\rm{total}}$ after $k$ iterates. This corresponds to a decay in the signal, i.e.
\begin{align}
    e^{-\lambda} = (1-r)^{ND},
    \label{eq:lamb_conversion}
\end{align} 
which makes the exponential decay parameter $\lambda$ that we introduced in Eq.~\eqref{eq:algorithmic_noise_model} to be $\lambda = -\ln((1-r)^{ND})$. 
Similarly, we can substitute $r$ from Eq.~\eqref{eq:single_qubit_depolarizing} into the variance expression $\rm{Var}[\langle P\rangle]$ from Eq.~\eqref{eq:quantum_var}
\begin{align}
    \text{Var}[\langle P\rangle]=\frac{1}{2^N+1}\big[\big((1-r)^2+3(r/3&)^2\big)^{NDk}\nonumber \\ 
    &-(1-r)^{2NDk}\big].
\end{align}
We note that $\text{Var}[\langle P\rangle]$ here establishes the expected deviation $\beta_j$ on our signal $f_j$ in our algorithmic error model to be discussed next.

\begin{figure}[!htb]
\includegraphics[width=0.48\textwidth]{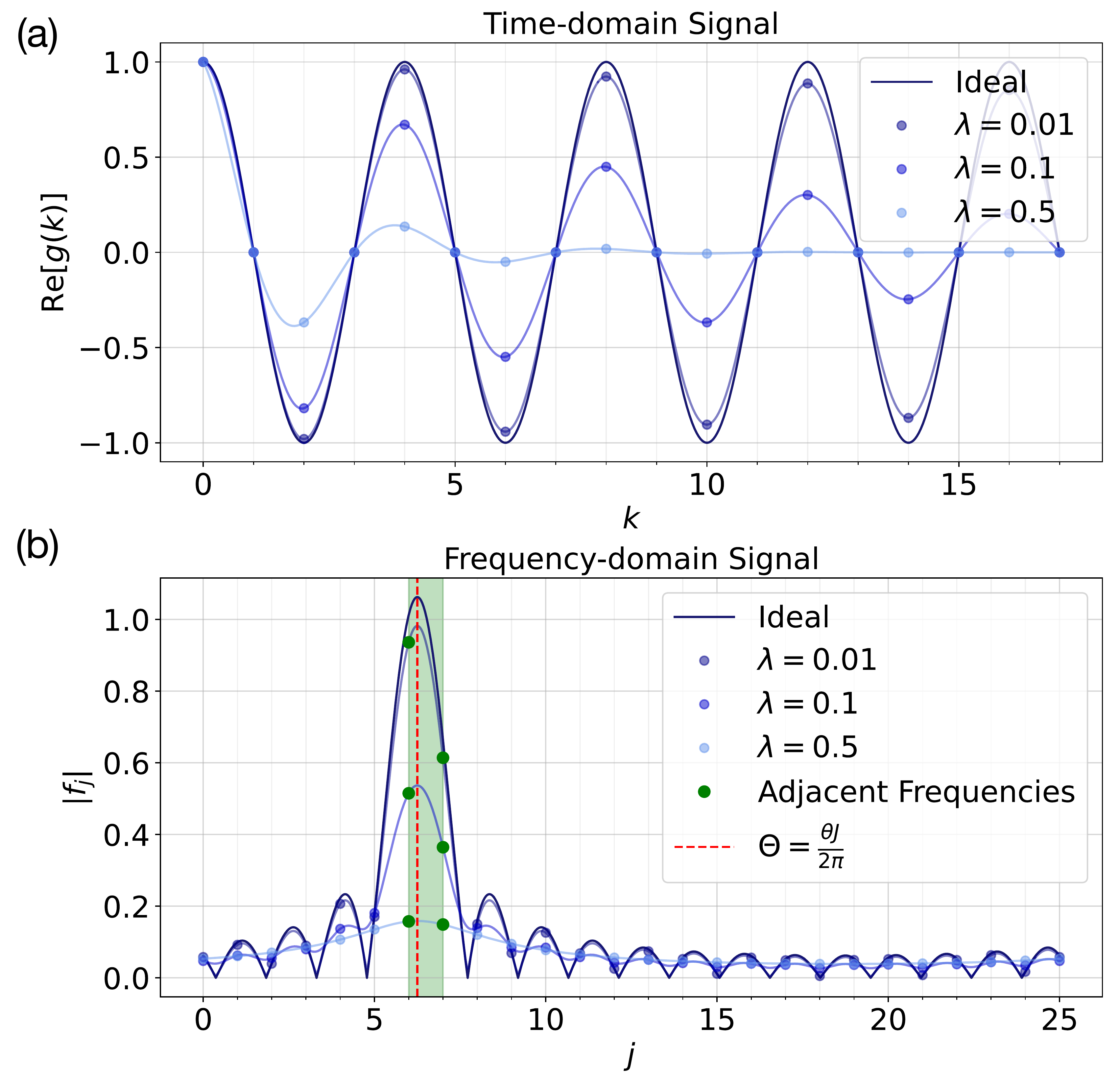}
\caption{Expected $\textup{Re}[g(k)]$ \textbf{(a)} and $|f_j|$ \textbf{(b)} under the exponential decay noise with decay parameter $\lambda=0.01,0.1,0.5$.}
\label{fig:efj_vs_j_noisy}
\end{figure}

Under the effects of random fluctuations and exponential decay, we now arrive at a noisy expression of the expected $\hat{f}_j(k,\phi,z)$, similar to the calculation in Eq.~\eqref{eq:expect_noiseless}
\begin{align}
\label{eq:noisy_exval}
\mathbb{E}[\hat{f}_j(k,\phi,z)]=\beta_j+\frac{1}{K}\left(\frac{1-e^{K(i\theta-i2\pi j/J-\lambda)}}{1-e^{i\theta-i2\pi j/J-\lambda}}\right), 
\end{align}
where $\beta_j = \mathbb{E}_k[\eta_{j,k}]$ is the expected deviation of our signal related to the quantity $\text{Var}[\langle P\rangle]$. We delay the detailed discussion of the variance to Appendix \S\ref{app:variance}. There, we numerically show that the standard deviation of $\langle P\rangle$,  $\sigma_{\langle P\rangle}:=\sqrt{\rm{Var}[\langle P\rangle]}$, is negligible compared to the signal magnitude $\sim 1$ for the parameter settings of our interest in the EFTQC regime. This means that the variance term drops out of our algorithmic error model~{\eqref{eq:algorithmic_noise_model}}, leaving exponential decay on the ideal signal as the sole source of error to be considered for the remaining analysis. This result is indicative that our approach for modeling the noisy trajectory states with a unitary 2-design model is indeed consistent with a global depolarizing noise, the effect of which has been thoroughly studied in many EFTQC algorithms~\mbox{\cite{obrien_quantum_2019, katabarwa2021reducing, ding2023robust}}. Hence, we set the $\beta_j$ term to 0 for the rest of the discussion. 

The effect of exponential decay on the expected signal from Eq.~\eqref{eq:noisy_exval} is shown as a function of decay strength $\lambda$ in Fig.~\ref{fig:efj_vs_j_noisy}. We note that as $\lambda$ increases, the signal $g(k)$ decays exponentially as a function of circuit depth $k$, which results in a flatter spectrum in the expected Fourier signal amplitude $|f_j|$. This is problematic as the RFE algorithm relies on distinguishing the largest peak in the signal amplitude as our predicted phase index. A flattened spectrum means that the amplitude contrasts between neighboring peaks will decrease, and as a result, more measurements are needed to overcome shot noise in order to better discern the highest-amplitude point in the spectrum.

%% file: v2/5_Performance.tex
\subsection{Link to Algorithm Performance Analysis}
\label{subsec:algperformance}

The goal of this section is to establish the connection between the algorithmic noise model proposed in the previous section and the algorithm performance guarantee. 
As a reminder, the algorithm is said to succeed when the estimate $\hat{\theta}$ is within $\epsilon$ of $\theta$, see Fig.~\ref{fig:noiseless_rfe}(b).
Our analysis determines an upper bound on the number of samples that are needed to ensure success with probability greater than $1-\delta$.

The estimated $\hat{\theta}$ is calculated based on the discrete frequency point $2\pi j/J$ corresponding to the highest amplitude of $|\hat{f}_j|$ (as in Eq.~\eqref{eq:theta_estimate}).
As a reminder, the consideration of a successful estimate of $\theta$ depends on whether $\theta$ is one of the values of $2\pi j/J$ or not.
In the case when $\theta$ falls onto one of the discrete frequency points, there are a total of three values of $j$ leading to successful estimates, i.e., $j= J\theta /2\pi, J\theta /2\pi\pm1$. 
In the case when $\theta$ falls between two points in the discrete frequency spectrum, as shown in Fig.~\ref{fig:efj_vs_j_noisy}(b), there are a total of two neighboring points that constitute successful guesses, namely $\lfloor J\theta/2\pi\rfloor$ and $\lceil J\theta/2\pi\rceil$.

Let $j_*$ indicate the index of the ``best estimate,'' that is, the integer closest to $J\theta/2\pi$.  
Any index $j$ that is more than 1 away from $J\theta/2\pi$ will not correspond to an $\epsilon$-accurate estimate according to our criteria for success.
We refer to such estimates as ``bad estimates.''
Hence, to guarantee that the algorithm succeeds, it suffices that
$|\hat{f}_j|^2< |\hat{f}_{j*}|^2$ for all $j$ with $|j-J\theta/2\pi|>1$. In other words, this condition ensures that no bad estimate has the largest $|f_j|$ so that the algorithm picks one of the $\epsilon$-accurate estimates.
Note that this is not a necessary condition; the algorithm could succeed even if $|\hat{f}_{j*}|^2$ were smaller than that of one of the bad estimates, as long as the other (or one among the two others) $\epsilon$-accurate estimate had a magnitude larger than the rest.

The above algorithm success condition can be violated in the presence of exponential decay noise, the effect of which is shown in Fig.~\ref{fig:efj_vs_j_noisy}(b). 
The noise reduces contrast between neighboring peaks, which can invalidate our success condition if not enough measurement samples are collected to overcome the uncertainty due to shot noise. One way to counteract this deleterious effect is to perform extra measurements. Thus the performance of our algorithm can be quantified by establishing an upper bound on the number of measurements needed to guarantee a certain success probability of the algorithm, for various noise strengths and the desired accuracies. We point interested readers to Appendix \S\ref{app:algorithm_performance_derivation} for a proof of the algorithm performance bound. 

\begin{figure}
\includegraphics[width=0.48\textwidth]{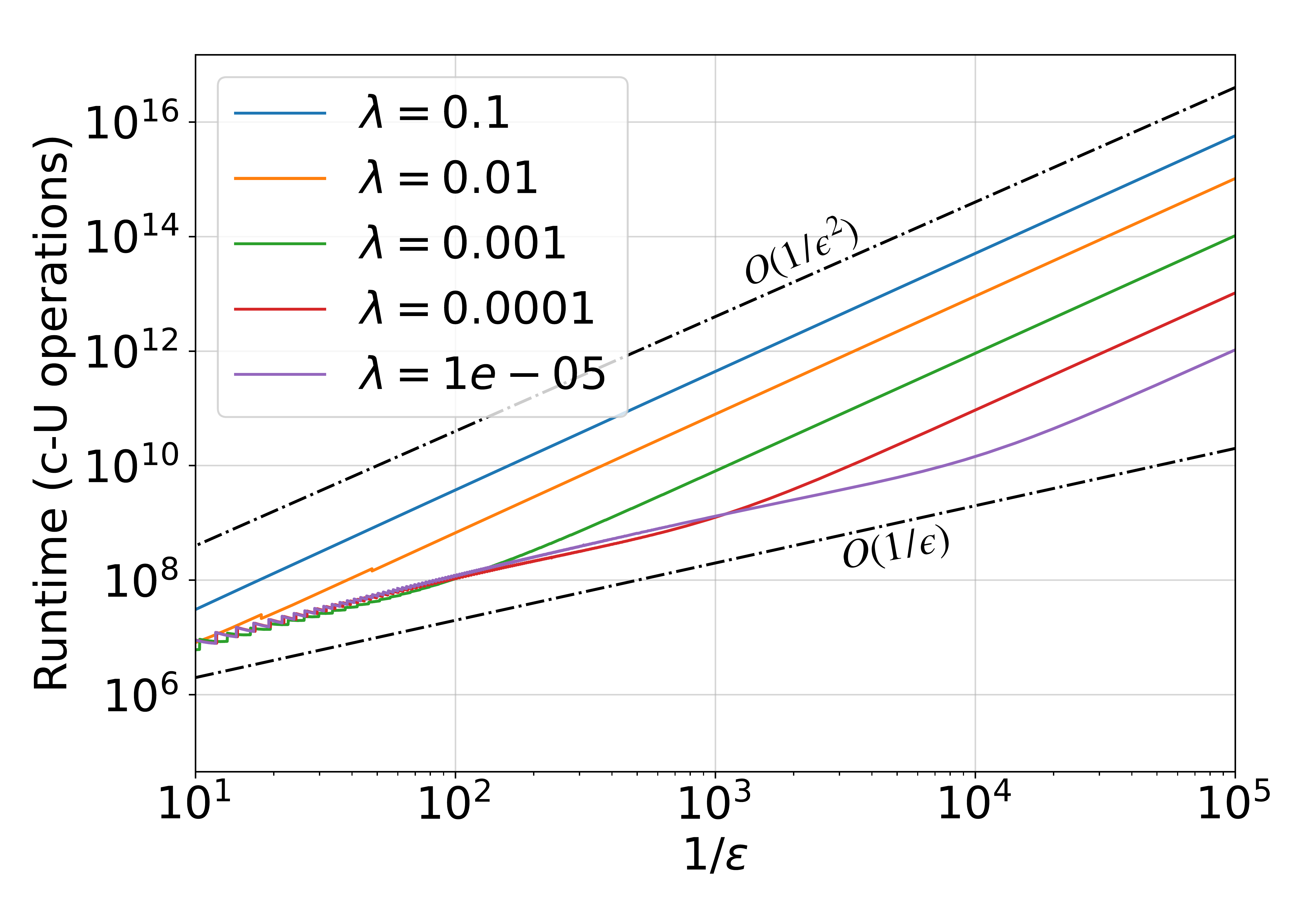}
    \caption{This plot shows the runtime (measured in total number of calls to the $c$-$U$ operation) as a function of the target accuracy inverse $1/\epsilon$ in the presence of various exponential decay errors of strength $\lambda=0.1, 0.01, 0.001, 0.0001$, and $ 0.00001$.
    As $\lambda$ decreases, the runtime transitions from a $1/\epsilon^2$ scaling to a $1/\epsilon$ scaling. 
    The upper and lower dotted lines show the $O(1/\epsilon^2)$ and $O(1/\epsilon)$ scaling, respectively.}
\label{fig:rainbow}
\end{figure}

We now present the main result of our algorithm performance analysis as follows. 
For target accuracy $\epsilon$ and exponential decay parameter $\lambda$,
a success probability greater than $1-\delta$ can be ensured by using $M$ samples satisfying \begin{align}
    M\geq 8W(K(\epsilon, \lambda),2\pi/\epsilon,\lambda)\log(16\pi/\delta\epsilon),
    \label{eq:perf_bound}
\end{align}
where $W(K,J,\lambda)$ is a complicated function whose explicit form is described in Eq.~\eqref{eq:explicit_M_bound} of 
Appendix \S\ref{app:algorithm_performance_derivation} and $K$ is chosen as a function of $\epsilon$ and $\lambda$ as described below.
The algorithm performance measured by the runtime upper bound in units of $c$-$U$ operations are plotted against $1/\epsilon$ for various values of $\lambda$ in Fig.~\ref{fig:rainbow}. In the low-noise regime, the runtime scales as $1/\epsilon$, which resembles that of the original QPE algorithm \cite{qpe}. In the high-noise regime, the runtime scales as $1/\epsilon^2$. For a moderate amount of noise, the runtime performance of the algorithm interpolates between the low- and high-noise performance. 

Due to the limited coherence time from the exponential decay, we choose the maximal circuit depth $K$ based on $\epsilon$ and $\lambda$, given by $K(\epsilon,\lambda)=\max\{c\lfloor\frac{1}{c(a\lambda+\epsilon b/2\pi)}\rfloor,2\}$, where $a,b,c\in \mathbb{R}^+$. 
We note that here we set $K$ to be tunable based on $\lambda$ and $\epsilon$ such that the runtime of the algorithm interpolates between the Heisenberg-limit scaling (i.e. $O(1/\epsilon)$) in the regime of $J\lambda\gg 1$ and the shot-noise-limit scaling (i.e. $O(1/\epsilon^2)$) in the regime of $J\lambda\ll 1$.
For the specific functional form of $K$, we employ the floor function and set $c=10$ to bypass values of $K$ between 3 and 9, which were numerically found to be suboptimal. The parameters $a=2$ and $b=1.5$ were then chosen within this form to roughly minimize the runtime upper bound over a range of values of $J$ and $\lambda$. 
We note that since this functional form of $K$ is empirically derived to minimize the runtime upper bound, in practice, a more rigorous treatment for developing an optimal strategy for choosing $K$ is necessary in the future.

Our analysis of the algorithm in Appendix \S\ref{app:algorithm_performance_derivation} differs from that of \cite{kshirsagar2022proving} in two regards: 1) we separate the roles of $K$ and $J$ enabling high-accuracy estimates 
in the high-noise setting
and 2) we take into consideration the correlation between the values of neighboring $\hat{f}_j$, enabling a reduction in sample complexity in the high-noise (low-$K$) setting.
These two features enable us to establish a unifying expression that captures the performance of the algorithm in a wide variety of scenarios, ranging from using (effectively) a Bernoulli estimation approach (i.e. $K=2$) to a Heisenberg-limited phase estimation approach (i.e. $K=O(1/\epsilon)$). A signature of the switch in the algorithm's strategy from the Bernoulli to the Heisenberg approach to accommodate different scenarios of error is marked by the cusp in the runtime upper bound of RFE in Fig.~\ref{fig:ft_overhead}.

Finally, we would like to address the issue of determining the exponential decay parameter $\lambda$. Our algorithm relies on some knowledge of $\lambda$ when setting the appropriate $K(\epsilon, \lambda)$.
This raises the question of whether it is necessary to accurately determine $\lambda$ and, if not, to what extent a discrepancy between the presumed and actual values of $\lambda$ would compromise the algorithm's performance.
The precise level of accuracy required to maintain the derived runtime remains an open question that we defer to future investigation. Nonetheless, intuitively, we anticipate that overestimating $\lambda$ will result in using more samples than necessary, while underestimating $\lambda$ will lead to using fewer samples than required, thus slightly increasing the probability of failure. We further highlight that several established benchmarking techniques, such as randomized benchmarking \cite{RB} and cross-entropy benchmarking \cite{XEB}, can be utilized to estimate essential noise parameters like the depolarizing error rate. These estimates can then be applied to determine the value of $\lambda$ using Eq.~\eqref{eq:lamb_conversion} in our methodology.

\subsection{FT Overhead Estimation Comparison}
\label{subsec:ft_overhead}

\begin{figure}[!htb]
\includegraphics[width=0.49\textwidth]{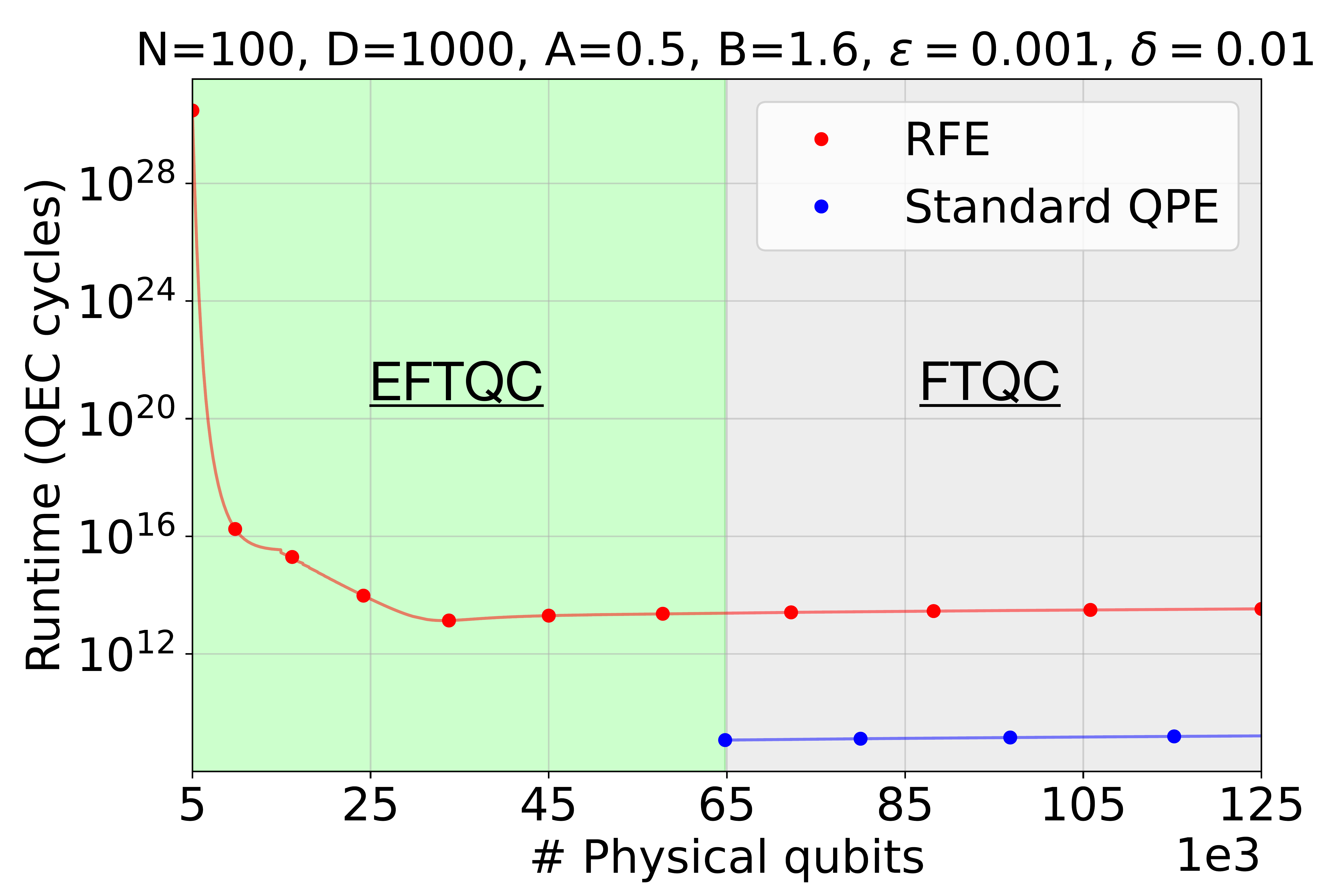}
    \caption{Comparison of runtime upper bounds (in units of error correction cycles) between RFE (red) and the traditional QPE algorithm (blue) as a function of physical qubits available. These upper bounds are proved in Appendix Sections {\ref{app:algorithm_performance_derivation}} and {\ref{app:qpe_ft_deriavtion}}, respectively. The upper bound on the runtime of RFE may be higher than the standard approach to QPE, RFE can be run using an order-of-magnitude fewer physical qubits. The cusp occurring in the runtime upper bound of RFE at $\sim 10^4$ physical qubits marks the switch of the algorithm's strategy from using $K=2$ to $K>2$. In the regime of low physical qubits, the runtime becomes extremely high due to the sampling overhead that compensates for the high degree of error. Note that the runtime of standard QPE may be larger than presented; we've optimistically assumed that a single sample of the QPE circuit suffices, though in practice one might have to take several samples to ensure the failure rate is below some tolerance.}
\label{fig:ft_overhead}
\end{figure}

As an application of our established performance bound in Eq.~\eqref{eq:perf_bound}, in this section we provide a cost analysis of implementing the RFE algorithm accounting for the error correction overhead. Specifically, we aim to compare the overall runtime performance (measured in units of QEC cycles) of the traditional QPE versus the RFE algorithm as a function of the number of physical qubits required, which relates to distance of the surface code $d$ as described in Eq.~\eqref{eq:FT_model}. 

In Fig.~\ref{fig:ft_overhead}, we show this comparison for the case of a logical $c$-$U$ acting on 100 qubits with a unitary circuit depth 1000, aiming for an accuracy of $0.1\%$ and failure probability below $1\%$. One notable characteristic of the traditional QPE algorithm is that there is a minimal number of physical qubits/code distance below which the desired algorithm success probability cannot be achieved (refer to the detailed derivation in Appendix \S\ref{app:qpe_ft_deriavtion}). Therefore, in order to implement the traditional QPE algorithm reliably, a certain distance code or number of physical qubits have to be attained. This is, however, not required by the RFE algorithm. In principle, a higher runtime cost can always be paid in exchange for fewer physical qubits or a lower code distance in implementing the RFE algorithm. This is particularly valuable in the EFTQC regime, where devices have limited number of physical qubits, as depicted schematically in the green region of Fig.~\ref{fig:ft_overhead}. 

The RFE algorithm offers the advantage of requiring an order-of-magnitude fewer qubits compared to the traditional QPE algorithm.
However, Fig.~\ref{fig:ft_overhead} also shows that the runtime upper bound of RFE in the FTQC regime is approximately four orders of magnitude larger than that of traditional QPE. Nevertheless, we expect that the aforementioned upper bound is conservative for two main reasons. Firstly, the analysis incorporates several analytical bounds that result in a more cautious choice of $M$ than what is likely necessary (a well-known phenomenon in algorithm analysis). 
We expect that the empirical runtime would be several orders of magnitude smaller than the derived upper bounds and leave an investigation of this to future work.
Secondly, the algorithm itself can be optimized to enhance performance. Examples of potential improvements include: 1) tailoring the distribution from which $k$ is sampled based on the noise characteristics of the device, and 2) employing more sophisticated fitting strategies for extracting $\hat{\theta}$ from the Fourier transformed data.

%% file: v2/6_Conclusion.tex
In summary, we have developed a methodology for systematically analyzing the performance of a class of quantum algorithms suited for early fault-tolerant quantum computers.
This is motivated by the need to understand the quantum resources necessary for these algorithms to achieve quantum advantage.
Our approach can be extended to encompass various error models, fault-tolerant resource overhead models, and a wide range of quantum algorithms. By offering a generalized framework, our methodology paves the way for a comprehensive understanding of resource requirements and performance trade-offs in the realm of early fault-tolerant quantum computing.

As an application of our methodology, we analyzed the performance of the recently proposed RFE algorithm~\cite{kshirsagar2022proving}.
Studying the circuit-level error under our proposed Haar trajectory model, we found that the noise can be best described by an exponential decay at the algorithmic level. We developed a variant of this algorithm that interpolates between low-depth and high-depth (based on the strength of the exponential decay) and analytically derived its runtime upper bound as a function of target accuracy and failure rate.
Studying the algorithm under a continuum of noise strengths, we found that the runtime upper bound interpolates between $O(1/\epsilon^2)$ (shot-noise limited scaling) and $O(1/\epsilon)$ (Heisenberg scaling) in the high- to low-noise limits. 

Based on the runtime upper bound, we then carried out a fault-tolerant overhead
analysis, comparing RFE with the traditional QPE in the problem instance of 100 logical qubits. Our analysis showed that the RFE algorithm can be implemented with an order-of-magnitude fewer physical qubits than the traditional QPE algorithm, albeit with a substantial increase in runtime. This tradeoff allows for an earlier onset of practical quantum advantage in the EFTQC regime, where error correction remains expensive.

There are several crucial research directions that hold promise for delivering quantum computation with practical advantage:

First and foremost, it is imperative to develop more realistic circuit-level error models. Our methodology relies on the assumption that the sampled Haar trajectory states are derived from a unitary two-design, or at the very least, their statistics can be well-approximated by those of a two-design. To better understand the limitations of this model, further numerical investigations are required. In future studies, it would be valuable to develop an empirically-derived model that captures the expected bias and variance of various circuit-level errors. We anticipate that the structure of the specific quantum circuits and error models of interest will introduce non-uniformity in the distribution of noisy trajectories. This non-uniformity could in principle shift $\mathbb{E}[\langle P\rangle]$ and increase $\text{Var}[\langle P\rangle]$, leading to a worse algorithm performance than that predicted by the current analysis. By addressing these aspects, future studies can provide more realistic noise settings and evaluate the performance of Randomized Fourier Estimation (RFE) under those settings.

Secondly, there is a need to develop fault-tolerant overhead models specifically tailored for early fault-tolerant quantum computers. 
Recent work~\cite{fellous2021limitations} has shown that small deviations of fault-tolerant architectures from the ideal assumptions can substantially impact the performance of quantum algorithms.
We anticipate that by including such realistic models of fault-tolerant architectures into the methodology of this work, the importance of robust quantum algorithms will be increasingly apparent.

Thirdly, as briefly mentioned in \S{\ref{subsec:logical_error_model}}, we have exclusively considered circuit-level errors while neglecting the effect of state preparation errors in our noise modeling. Extending our methodology to include state preparation errors is a plausible avenue for further exploration and would likely require an algorithm-specific analysis. We note that parallel to our work, some related recent studies (e.g. {\cite{ni2023lowdepth}}) provide complementary insights by studying the impact of state preparation errors on the performance of a similar-spirited EFTQC algorithm to RFE. A potential future direction is to study the expected performance of these relatively shallow-circuit algorithms under a mixture of the realistic error models, which early fault-tolerant devices are known to suffer from. 

Furthermore, it is crucial to extend our methodology to analyze alternative EFTQC algorithms rather than RFE towards solving practical problems. We have chosen to analyze the RFE algorithm as a proof-of-principle case study to demonstrate the viability of our methodology, for the RFE is analytically tractable, which comes at the price of a potential sub-optimal performance. We note that, however, our methodology is readily applicable to more advanced versions of EFTQC algorithms, such as the QCELS algorithm, a variant of the phase estimation algorithm proposed in~{\cite{ding2023qcels}}. For example, the noise modeling and FT overhead estimation detailed in our paper can be directly transferable to the QCELS robustness analysis similar to that outlined in {\cite{ding2023robust}}. In addition, our noise modeling scheme allows for the substitution of depolarizing errors with alternative structured noise models, such as dephasing errors, with simple modifications to the statistics of the error channels modeled in {\S\ref{sec:noise_model}}. We expect that these alternative algorithms will also be found to reduce quantum resources in the early fault-tolerant regime.

We envision that by extending our methodology to encompass existing algorithms and future developments in the EFTQC regime, we can enable their comprehensive evaluation and enhance the practical utilization of early fault-tolerant quantum computers.



%% file: Appendix.tex
\section{Statistics of quantum expectation value} 
\label{app:stats}
\subsection{Expectation value $\mathbb{E}[\langle P\rangle]$}
For a generic $N$-qubit Pauli operator $P$, the expected value of the quantum expectation value $\langle P\rangle$ over all possible Haar trajectory states sampled from a spherical 2-design is given by
\begin{align}
    \mathbb{E}[\langle P\rangle] &=p_{\mathbf{0}}\mathbb{E}[\Tr[P|\psi_{\mathbf{0}}\rangle\langle\psi_{\mathbf{0}}|]]+\sum_{\mathbf{j}=1}^{4^{ND}-1}p_{\mathbf{j}}\mathbb{E}[\Tr[P|\psi_{\mathbf{j}}\rangle\langle\psi_{\mathbf{j}}|]]\\
    &= p_{\mathbf{0}}\langle\psi_{\mathbf{0}}|P|\psi_{\mathbf{0}}\rangle+\sum_{\mathbf{j}=1}^{4^{ND}-1}p_{\mathbf{j}}\bigg(\frac{1}{|S^{2^N}|}\int_{\psi_{\mathbf{j}}\in S^{2^N}}\Tr[P|\psi_{\mathbf{j}}\rangle\langle\psi_{\mathbf{j}}|] d\psi_{\mathbf{j}}\bigg)\\
    &= p_{\mathbf{0}}\langle\psi_{\mathbf{0}}|P|\psi_{\mathbf{0}}\rangle+\sum_{\mathbf{j}=1}^{4^{ND}-1}p_{\mathbf{j}}\bigg(\Tr[P\frac{1}{|S^{2^N}|}\int_{\psi_{\mathbf{j}}\in S^{2^N}}|\psi_{\mathbf{j}}\rangle\langle\psi_{\mathbf{j}}|d\psi_{\mathbf{j}}] \bigg).
\end{align}
Here we have distributed the integral within the $\Tr$ operator due to its linearity and arrived at the expression of an average $N$-qubit Haar state
\begin{align}
    \frac{1}{|S^{2^N}|}\int_{\psi_{\mathbf{j}}\in S^{2^N}}|\psi_{\mathbf{j}}\rangle\langle\psi_{\mathbf{j}}|d\psi_{\mathbf{j}} = I/2^{N}.
\end{align}
It follows that the expectation value is
\begin{align}
    \mathbb{E}[\langle P\rangle]&= p_{\mathbf{0}}\langle\psi_{\mathbf{0}}|P|\psi_{\mathbf{0}}\rangle+\sum_{\mathbf{j}=1}^{4^{ND}-1}p_{\mathbf{j}}\Tr[PI/2^N]\\
    &= p_{\mathbf{0}}\langle\psi_{\mathbf{0}}|P|\psi_{\mathbf{0}}\rangle+\sum_{\mathbf{j}=1}^{4^{ND}-1}p_{\mathbf{j}}/2^N\Tr[P]\\
    &= p_{\mathbf{0}}\langle\psi_{\mathbf{0}}|P|\psi_{\mathbf{0}}\rangle.
    \label{eq:quantum_exp_exp_sup}
\end{align}
\subsection{Variance $\text{Var}[\langle P\rangle]$}
\label{app:variance}
Similarly, we can compute the variance of $\langle P\rangle$
\begin{align}
    \text{Var}[\langle P\rangle] = \mathbb{E}[\langle P\rangle^2]-\mathbb{E}[\langle P\rangle]^2 = \mathbb{E}[\langle P\rangle^2]-p_{\mathbf{0}}^2\langle\psi_{\mathbf{0}}|P|\psi_{\mathbf{0}}\rangle^2,
\end{align}
where
\begin{align}
    \mathbb{E}[\langle P\rangle^2] &= \sum_{\mathbf{i},\mathbf{j}=0}^{4^{ND}-1}p_{\mathbf{i}}p_{\mathbf{j}}\mathbb{E}[\Tr[P|\psi_{\mathbf{i}}\rangle\langle\psi_{\mathbf{i}}|]\Tr[P|\psi_\mathbf{j}\rangle\langle\psi_\mathbf{j}|]]\\
    &= \sum_{\mathbf{i},\mathbf{j}=0}^{4^{ND}-1}p_{\mathbf{i}}p_{\mathbf{j}}\bigg(\frac{1}{|S^{2^N}|^2}\int_{\psi_{\mathbf{i}}\in S^{2^N}}\int_{\psi_\mathbf{j}\in S^{2^N}}\Tr[P|\psi_{\mathbf{i}}\rangle\langle\psi_{\mathbf{i}}|]\Tr[P|\psi_\mathbf{j}\rangle\langle\psi_\mathbf{j}|]d\psi_{\mathbf{i}}d\psi_\mathbf{j}\bigg).
    \label{eq:29}
\end{align}
We note that here the summation of cross-terms where $\mathbf{i}\neq \mathbf{j}$ within Eq.~\eqref{eq:29} evaluates to 0.
Summing the rest of the terms where $\mathbf{i}=\mathbf{j}$, excluding the noiseless term $p_{\mathbf{0}}^2\langle\psi_{\mathbf{0}}|P|\psi_{\mathbf{0}}\rangle^2$, we get
\begin{align}
    \text{Var}[\langle P\rangle]&=\sum_{\mathbf{j}=1}^{4^{ND-1}}p_{\mathbf{j}}^2\mathbb{E}[\Tr[P|\psi_\mathbf{j}\rangle\langle\psi_\mathbf{j}|]^2]\\
    &=\sum_{\mathbf{j}=1}^{4^{ND}-1}p_{\mathbf{j}}^2\bigg(\frac{1}{|S^{2^N}|}\int_{\psi_\mathbf{j}\in S^{2^N}}\Tr[(P\otimes P)|\psi_\mathbf{j}\rangle\langle\psi_\mathbf{j}|\otimes |\psi_\mathbf{j}\rangle\langle\psi_\mathbf{j}|]d\psi_\mathbf{j}\bigg)\\
    &= \sum_{\mathbf{j}=1}^{4^{ND}-1}p_{\mathbf{j}}^2\bigg(\Tr[(P\otimes P)\frac{1}{|S^{2^N}|}\int_{\psi_\mathbf{j}\in S^{2^N}}|\psi_\mathbf{j}\rangle\langle\psi_\mathbf{j}|\otimes |\psi_\mathbf{j}\rangle\langle\psi_\mathbf{j}|d\psi_\mathbf{j}]\bigg)\\
    &= \sum_{\mathbf{j}=1}^{4^{ND}-1}p_{\mathbf{j}}^2\Tr[P\otimes P\frac{I\otimes I +\text{SWAP}}{2\Tr[(I\otimes I+\text{SWAP})/2]}]\\
    &= \sum_{\mathbf{j}=1}^{4^{ND}-1}p_{\mathbf{j}}^2\Tr[\frac{P\otimes P +(P\otimes P)\text{SWAP}}{\Tr[I\otimes I+\text{SWAP}]}]\\
    &= \sum_{\mathbf{j}=1}^{4^{ND}-1}p_{\mathbf{j}}^2\bigg(\Tr[\frac{P\otimes P}{\Tr[I\otimes I+\text{SWAP}]}]+\Tr[\frac{(P\otimes P)\text{SWAP}}{\Tr[I\otimes I+\text{SWAP}]}]\bigg)\\
    &= \sum_{\mathbf{j}=1}^{4^{ND}-1}\frac{p_{\mathbf{j}}^2}{\Tr[I\otimes I+\text{SWAP}]}\Tr[(P\otimes P)\text{SWAP}]\\
    &= \sum_{\mathbf{j}=1}^{4^{ND}-1}\frac{p_{\mathbf{j}}^2}{\Tr[I\otimes I+\text{SWAP}]}\Tr[P^2]\\
    &=\frac{2^N}{\Tr[I\otimes I+\text{SWAP}]}\sum_{\mathbf{j}=1}^{4^{ND}-1}p_{\mathbf{j}}^2\\
    &=\frac{2^N}{\Tr[I\otimes I]+\Tr[\text{SWAP}]}\sum_{\mathbf{j}=1}^{4^{ND}-1}p_{\mathbf{j}}^2\\
    &=\frac{2^N}{2^{2N}+2^N}\sum_{\mathbf{j}=1}^{4^{ND}-1}p_{\mathbf{j}}^2\\
    &=\frac{1}{2^{N}+1}\sum_{\mathbf{j}=1}^{4^{ND}-1}p_{\mathbf{j}}^2.
\end{align}

For the single-qubit depolarizing error channel proposed in Eq.~\eqref{eq:single_qubit_depolarizing} with error probability $r$, the variance expression becomes
\begin{align}
    \text{Var}[\langle P\rangle] 
    & = \frac{1}{2^{N}+1}\left[\left(\sum_{\mathbf{j}=0}^{4^{ND}-1}p_{\mathbf{j}}^2 \right)- p_{\mathbf{0}}^2 \right] \\
    &=\frac{1}{2^N+1}\left[\left((1-r)^2+3(r/3)^2\right)^{NDk}-(1-r)^{2NDk}\right].
\end{align}

\begin{figure}
\includegraphics[width=\textwidth]{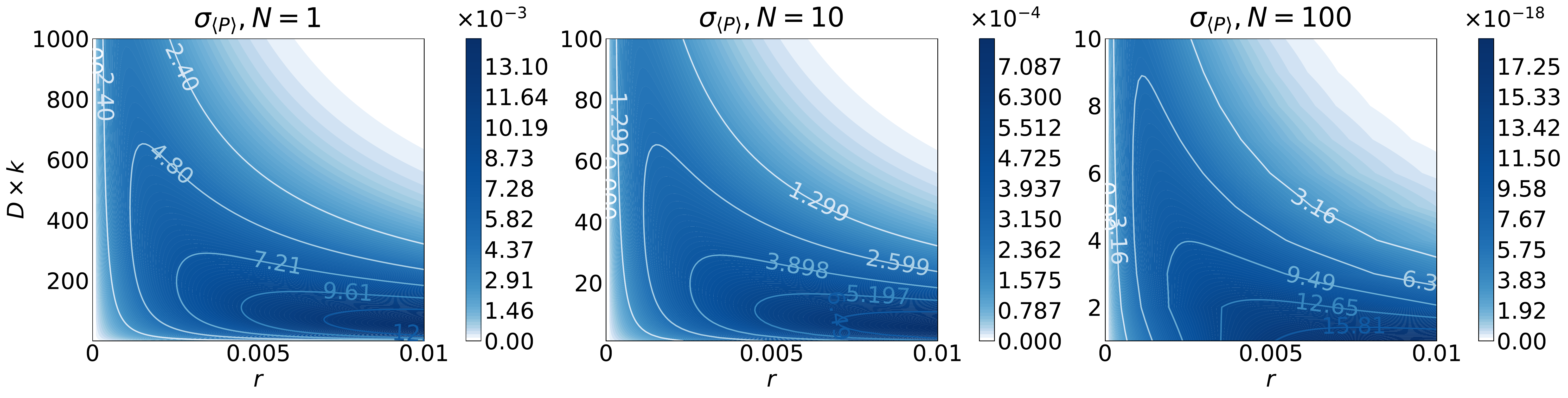}
    \caption{Standard deviation of $\langle P \rangle$ as a function of $D\times k$ and $r$ is plotted on a log scale for $N=1,10,100$.} 
\label{fig:std_p}
\end{figure}

Fig.~\ref{fig:std_p} illustrates the standard deviation of $\langle P \rangle$, $\sigma_{\langle P \rangle}:=\sqrt{\rm{Var}[\langle P \rangle]}$ as a function of depths $D\times k$ and single-qubit depolarizing rate $r$, considering number of logical qubits $N=1,10,100$. We observe that $\sigma_{\langle P \rangle}$ remains relatively small overall for $N=100$, while it is comparatively larger for $N=1$ and 10. We note that $\sigma_{\langle P \rangle}$ is particularly large for low-depth and high-error settings. In the extreme case of $N=1$, $\sigma_{\langle P \rangle}$ reaches a concerning high magnitude $\sim 10^{-2}$, which is too large to be overlooked in our algorithmic error model. 

Nonetheless, we opted to exclude the random fluctuation term stemming from variance in our algorithmic error model throughout our analysis. This decision aligns with the objective of our study, which focuses on examining the performance of devices in the early fault-tolerant regime. Specifically, we are interested in the parameter regime characterized by $r=p_{\rm{logical}}\lesssim 10^{-2}$, $N\gtrsim 100$, and $D\times k \gtrsim 1000$. In this regime, the standard deviation $\sigma_{\langle P \rangle}$ remains sufficiently small, justifying our assumption to omit it from the algorithmic error model. We point out that the performance of the RFE algorithm under the NISQ setting, where $p_{\rm{logical}} > 10^{-2}$, $N < 100$, and $D\times k < 1000$, falls outside the scope of our study and merits further investigation.

\section{Algorithm Performance}
\label{app:algorithm_performance_derivation}

In this section, we detail the analytical derivation of the performance upper bound of the RFE algorithm in the presence of exponential decay noise with strength $\lambda$. Rather than presenting these results in a theorem-proof format, we choose a more narrative presentation to facilitate broader accessibility. Based on the success condition $|\hat{f}_j|^2< |\hat{f}_{j*}|^2$ for all $j$ with $|j-\Theta|>1$, where $\Theta=J\theta/2\pi$, as discussed in \S\ref{subsec:algperformance}, the algorithm failure probability can be upper bounded using the union bound
\begin{align}
    \textup{Pr}(\textup{fail})&\leq \sum_{|j-\Theta|\geq 1}\textup{Pr}(\textup{fail}_j)\\
    &\leq (J-2)\max_j\textup{Pr}(\textup{fail}_{j})\\
    &\leq J\max_j\textup{Pr}(\textup{fail}_{j}).
\end{align}
We will upper bound this worst-case failure probability using Chebyshev's inequality.


In previous analyses of the RFE algorithm \cite{kshirsagar2022proving}, we had analyzed the likelihood that every Fourier estimate $\hat{f}_j$ was within some distance of its mean $f_j$.
This approach does not capture the performance of the algorithm in the regime of small $K$, as the correlation between Fourier estimates, $\hat{f}_j$ and $\hat{f}_{j'}$, is largely overlooked.
As a result, the success probability bound becomes too loose to capture the actual performance scaling of the algorithm.

To account for the correlation among the Fourier estimates, we will consider a change of variables,
\begin{align}
    \hat{c}_{j,j*}& \equiv \hat{f}_j + \hat{f}_{j*}\\    
    \hat{d}_{j,j*}& \equiv \hat{f}_j - \hat{f}_{j*}.    
\end{align}
The motivation for this choice is the fact that $\hat{d}_{j,j*}$ captures correlation between $\hat{f}_j$ and $\hat{f}_{j*}$ and therefore will have small variance for nearby $j,j_*$ when $K$ is small compared to $J$.
In contrast, ignoring the correlation between  $\hat{f}_j$ and $\hat{f}_{j*}$ and 
treating their variances
separately, as was done in \cite{kshirsagar2022proving}, leads to an overestimation of the algorithm failure probability.
We will define the expectation values of these quantities to be 
\begin{align}
    c_{j,j'}& \equiv f_j + f_{j'}\\    
    d_{j,j'}& \equiv f_j - f_{j'}.    
\end{align}
The sufficient condition for success can be expressed as
\begin{align}
\label{eq:success_condition}
\textup{Re}(\hat{c}_{j,j*}   
\hat{d}_{j,j*}^*)<0 \textup{ for all } j \textup{ with }|j-J\theta/2\pi|>1
\end{align}
using the following relationship,
\begin{align}
|\hat{f}_j|^2-|\hat{f}_{j*}|^2&=\textup{Re}((\hat{f}_j + \hat{f}_{j*})(\hat{f}_j - \hat{f}_{j*})^*)\nonumber\\
&=\textup{Re}(\hat{c}_{j,j*}  
\hat{d}_{j,j*}^*).
\end{align}
This condition also holds for $c_{j,j*}$ and $d_{j,j*}$, which are the expected values of $\hat{c}_{j,j*}$ and $\hat{d}_{j,j*}$, respectively.
It is the statistical fluctuations from the finite sampling that can cause the condition to fail.
As the number of samples is increased, the estimates $\hat{c}_{j,j*}$ and $\hat{d}_{j,j*}$ will be expected to increasingly concentrate about their means.
We can quantify this concentration with the following 
version of the central limit theorem proven by H. Chernov~\cite{kothari2015large}, 
\begin{align}
    \textup{Pr}\left(\left|\frac{1}{M}\sum_{i=1}^M\hat{q}_i-\mathbb{E}\hat{q}\right|>t\right)\leq4 \exp(Mt^2/8\textup{Var}(\hat{q}))
\end{align}
where $\hat{q}$ is any continuous random complex variable.
We use the Chebyshev inequality to upper bound the likelihood that the estimator deviates from its \emph{ideal} expected value:
\begin{align}
\label{eq:chebyshev}
\textup{Pr}(|\hat{c}-\mathbb{E}\hat{c}|>\chi)\leq 4 \exp(-M\chi^2/8\textup{Var}(\hat{c}))\nonumber\\
\textup{Pr}(|\hat{d}-\mathbb{E}\hat{d}|>\eta)\leq 4 \exp(-M\eta^2/8\textup{Var}(\hat{d}))
\end{align}
where $\textup{Var}(\hat{c})$ and $\textup{Var}(\hat{d})$ are the single sample variances of $\hat{c}$ and $\hat{d}$.
Thus, we have two free parameters ($\chi$ and $\eta$) to choose in a manner such that 1)
$|\hat{c}-\mathbb{E}\hat{c}|\leq\chi$ and 
$|\hat{d}-\mathbb{E}\hat{d}|\leq\eta$ imply success (letting us upper bound the failure probability) and 2) the number of samples does not become too high.
While it does not necessarily minimize the number of samples, we will choose the value of $\chi$ such that the upper bounds in Eq.~\ref{eq:chebyshev} become equal: 
$\chi=\eta\sqrt{\textup{Var}(\hat{c})/\textup{Var}(\hat{d})}$.

Next, we aim to establish a sufficiently large value for $\eta$, while still ensuring that $|\hat{c}-\mathbb{E}\hat{c}|\leq\eta\sqrt{\textup{Var}(\hat{c})/\textup{Var}(\hat{d})}$ and 
$|\hat{d}-\mathbb{E}\hat{d}|\leq\eta$ imply success.
This will be achieved by recasting the success condition.
Note that the condition $\textup{Re}(\hat{c}_{j,j*}  
\hat{d}_{j,j*}^*)<0$ is independent of the magnitudes of $\hat{c}_{j,j*}$ and $\hat{d}_{j,j*}^*$; it only depends on the relative phase angle between these two complex numbers.
Thus, an equivalent condition for algorithm success is that, for all $j$ with $|j-J\theta/2\pi|>1$, the phase angle formed between complex values $\hat{c}_{j,j*}$ and $
\hat{d}_{j,j*}$ is not within $[-\pi/2,\pi/2]$.
For the time being, we will ease the notation by dropping the $j,j*$ subscripts.

We use this phase angle condition to establish a sufficient condition for success in terms of the allowable sizes of deviations from the mean.
The largest possible angle formed between $\hat{c}$ and $c$ is given by $\sin(\gamma)=|\hat{c} - c|/|c|$ and the largest possible angle formed between $\hat{d}$ and $d$ is given by $\sin(\tau)=|\hat{d} - d|/|d|$.
We also define the angle between $c$ and $d$ according to $|c| |d|\cos(\alpha) = \textup{Re}(c d^*)$.
With these definitions, the smallest possible phase angle formed between $\hat{c}$ and $\hat{d}$ is $\alpha-\gamma-\tau$, hence, the condition for success becomes
\begin{align}
    \pi/2< \alpha-\gamma-\tau.
\end{align}
We use the bounds of 
$x\leq\arcsin{x}\leq \pi x/2$ for $0\leq x \leq 1$
to establish a sufficient condition for success that is amenable to using the Chebyshev inequality.
First, we have that $\gamma + \tau \leq \frac{\pi}{2}(|\hat{c} - c|/|c|+|\hat{d} - d|/|d|)$.
Next we have that 
$\cos(\alpha)=-\sin(\alpha-\pi/2)$, and so $\arcsin(-\textup{Re}(cd^*)/|c||d|)=\alpha-\pi/2$.
Using the lower bound on $\arcsin$, we then have $-\textup{Re}(cd^*)/|c||d|\leq\alpha-\pi/2$.
From the chain of inequalities 
\begin{align}
    -\textup{Re}(cd^*)/|c||d|<\alpha-\pi/2 < \gamma+\tau
<\frac{\pi}{2}(|\hat{c} - c|/|c|+|\hat{d} - d|/|d|),
\end{align}
we can establish the following implication
\begin{align}
    \alpha-\pi/2 &< \gamma+\tau\nonumber\\
    &\Rightarrow\nonumber\\
    -\textup{Re}(cd^*)/|c||d|&<\frac{\pi}{2}(|\hat{c} - c|/|c|+|\hat{d} - d|/|d|)
\end{align}
The contrapositive of this statement gives our sufficient condition for algorithm success
\begin{align}
\label{eq:geometric_inequality}
    -\textup{Re}(c   
d^*)&>\frac{\pi}{2}(|c||\hat{d} - d|+|d||\hat{c} - c|)\nonumber\\
&\Rightarrow\nonumber\\
\pi/2&< \alpha-\gamma-\tau.
\end{align}

We can then set the maximal allowable value for $\eta$ according to Eq.~\eqref{eq:geometric_inequality}.
Defining $\rho = \sqrt{\textup{Var}(\hat{c})/\textup{Var}(\hat{d})}$, this is
\begin{align}
    \eta &= -\frac{2}{\pi}\frac{\textup{Re}(c   
d^*)}{|c|+|d|\rho}\nonumber\\
&= \frac{2}{\pi}\frac{|\hat{f}_{j*}|^2-|\hat{f}_j|^2}{|c|+|d|\rho}.
\label{eq:eta_max}
\end{align}

Putting this all together, the probability of failure is upper bounded by
\begin{align}
\textup{Pr}(\textup{fail})
&\leq J\max_j\textup{Pr}(\textup{fail}_{j})\\
&\leq J\max_j\textup{Pr}(|\hat{c}-c|>\chi \textup{ or }
|\hat{d}-d|>\eta)\\
&\leq J\max_j(\textup{Pr}(|\hat{c}-c|>\chi )+\textup{Pr}(
|\hat{d}-d|>\eta))\\
&\leq 8 J\exp(-M\eta_{j',j*}^2/8\textup{Var}(\hat{d}_{j',j*}))\\
\end{align}
where we've reintroduced the indices $j',j*$ for clarity and are using $j'$ to indicate the index that realizes the maximization.
Let $W(K, J, \lambda)$ be a parameterized upper bound 
on the quantity $\textup{Var}(\hat{d}_{j',j*})/\eta_{j',j*}^2$ that is independent of $\theta$, $j*$, and $j'$. 
To ensure that $P(\textup{fail})\leq\delta$, we can choose $M$ to be a value such that
\begin{align}
    8 J\exp(-M\eta_{j',j*}^2/8\textup{Var}(\hat{d}_{j',j*}))\le \delta\\
\end{align}
We can then solve for such a value of $M$ as
\begin{align}
   &  -M\eta_{j',j*}^2/8\textup{Var}(\hat{d}_{j',j*}) \le \log{(\delta/8J)}\\
   &M\ge \frac{8\textup{Var}(\hat{d}_{j',j*})}{\eta_{j',j*}^2}\log(8J/\delta)
\end{align}
To achieve a high probability of success, it is also sufficient to set $M$ to be greater than a value that is larger than the right-hand side above. Let $W(K,J,\lambda)$ be a function satisfying
\begin{align}
   8W(K,J,\lambda)\log(8J/\delta)\ge \frac{8\textup{Var}(\hat{d}_{j',j*})}{\eta_{j',j*}^2}\log(8J/\delta)
\end{align}
Then, setting
\begin{align}
   &M\ge 8W(K,J,\lambda)\log(8J/\delta)
\end{align}
is sufficient to ensure success with high probability.
As described in the main text, $K$ and $J$ are set as functions of $\lambda$ and $\epsilon$, given by
\begin{align}
    J&\leftarrow 2\pi/\epsilon\\
    K&\leftarrow \max\{\lfloor((1/\lambda)^{-1}+(2\pi/\epsilon)^{-1})^{-1}\rfloor,2\},
    \label{eq:Kvalue}
\end{align}
where we see that $K$ is a harmonic average of $J$ and $1/\lambda$ and $K\leq J$. 
This will make the number of measurements $M$ an implicit function of $\epsilon$ and $\lambda$.

We now establish a function $W$ that satisfies 
$W\geq \frac{\textup{Var}(\hat{d})}{\eta^2}$. From Eq.~\eqref{eq:eta_max}, we can establish the upper bound,
\begin{align}
\frac{\sqrt{\textup{Var}(\hat{d})}}{\eta}&= \frac{\pi}{2}\frac{\sigma_{\hat{d}}|c|+\sigma_{\hat{c}}|d|}{{|f_{j*}|^2-|f_j|^2}}\\
    &\leq 2\pi\frac{\sigma_{\hat{d}}+|d|}{{|f_{j*}|^2-|f_j|^2}},
\end{align}
where in the second line we have used the fact that both $\sigma_{\hat{c}}=\sqrt{\textup{Var}(\hat{c})}$ and $|c|$ are upper bounded by 4. This follows from the fact that for a single sample estimate, $\hat{c}$ has magnitude at most $4$ as it is the sum of two quantities that have magnitude at most $2$.
Since $\textup{Var}(\hat{d})=\mathbb{E}(|\hat{d}|^2)-d^2$ is positive, we have that $\mathbb{E}(|\hat{d}|^2)\geq \textup{Var}(\hat{d})$ and $\mathbb{E}(|\hat{d}|^2)\geq |d|^2$.
This lets us further upper bound $\sqrt{\frac{\textup{Var}(\hat{d})}{\eta^2}}$ as
\begin{align}
\label{eq:ratio_bound}
\frac{\sqrt{\textup{Var}(\hat{d})}}{\eta}&\leq 4\pi\frac{\sqrt{\mathbb{E}(|\hat{d}|^2)}}{{|f_{j*}|^2-|f_j|^2}}.
\end{align}
Using $\omega=e^{-i2\pi/J}$, $\mathbb{E}(|\hat{d}|^2)$ is computed to be 
\begin{align}
\mathbb{E}(|\hat{d}|^2)
&=4\mathbb{E}(\left|e^{-i\phi}z(\omega^{jk}- \omega^{j_*k})\right|^2)\\  &=8(1- \mathbb{E}\textup{Re}(\omega^{k(j_*-j)}))\\ 
    &=8- \frac{4}{K}\left(\frac{1-\omega^{K(j_*-j)}}{1-\omega^{j_*-j}}+\frac{1-\omega^{-K(j_*-j)}}{1-\omega^{-j_*+j}}\right)\\ 
    &=8- 8\cos(\pi(K-1)\frac{j_*-j}{J})\frac{\sin(\pi K\frac{j_*-j}{J})}{K\sin(\pi\frac{j_*-j}{J})}.
\end{align}
Defining $x=\pi(j_*-j)/J$,
we establish a function upper bounding the quantity above
\begin{align}
  \mathbb{E}(|\hat{d}|^2)
    &= 8\left(1- (\cos(Kx)\cos(x)+\sin(Kx)\sin(x))\frac{\sin(Kx)}{K\sin(x)}\right)\\
    &= 8\left(1- \frac{\sin(2Kx)}{2K\tan(x)}-\sin^2(Kx)/K\right)\\
    &\leq 8\left(1- \frac{\sin(2Kx)}{2K\tan(x)}\right)\\
    &\leq 8\left(1- \frac{\sin(2Kx)}{2K\tan(x)}\right)\\
            &\leq \frac{32}{3}\left(1- \exp(-4(Kx)^2/7)\right),
\end{align}
where the last inequality is    established by using numerical software and exploiting the fact that $K\geq2$.
Observe that this function is monotonically increasing in $x$. 
We rewrite $x=(\theta-2\pi j/J)/2-(\theta-2\pi j_*/J)/2$. Using the fact that $|\theta-2\pi j_*/J|\leq \pi/J$ and that the above function is monotonic in $x$, we obtain an upper bound by setting $|x|=\pi/J+|\theta-2\pi j/J|/2$,
\begin{align}
\label{eq:boundA}
  \mathbb{E}(|\hat{d}|^2)
  &\leq \frac{32}{3}\left(1- \exp(-\frac{4K^2}{7}(\pi/J+|\theta-2\pi j/J|/2)^2)\right)\equiv Q(K,J;j).
\end{align}
Next we lower bound the denominator of Eq.~\eqref{eq:ratio_bound}.
From Eq.~\eqref{eq:noisy_exval}, we calculate
\begin{align}
    |f_j|^2=\frac{e^{-(K-1)\lambda}}{K^2}\left(\frac{\cosh(K\lambda)-\cos(K(\theta-2\pi j/J))}{\cosh(\lambda)-\cos(\theta-2\pi j/J)}\right).
\end{align}
Defining $y_*=\theta-2\pi j*/J$, we have that $|y_*|\leq \epsilon/2 = \pi/J$. 
Note that the above function, for $j=j*$, is symmetric about $y_*=0$ and thus depends only on $|y_*|$.
Using the fact that $K\leq J$, on the range $0\leq|y_*|\leq\pi/J$, the above function is monotonically decreasing (as observed numerically).
Thus, it achieves its minimum at $|y_*|=\pi/J$.
From this we lower bound $|f_{j*}|^2$ as
\begin{align} 
\label{eq:boundB}
|f_{j*}|^2\geq\frac{e^{-(K-1)\lambda}}{K^2}\left(\frac{\cosh(K\lambda)-\cos(\pi K/J)}{\cosh(\lambda)-\cos(\pi/J)}\right)\equiv R(K,J,\lambda).
\end{align}
Lastly, we upper bound the square magnitude of the Fourier coefficient of the non-adjacent frequencies, $|f_j|^2$, where $|J\theta/2\pi -j|\geq1$.
To simplify the analysis, we define
\begin{align}
m(x;K,\lambda):=\frac{e^{-(K-1)\lambda}}{K^2}\left(\frac{\cosh(K\lambda)-\cos(x)}{\cosh(\lambda)-\cos(x/K)}\right),
\end{align}
such that $|f_j|^2=m(K(\theta-2\pi j/J); K,\lambda)$. 
We will establish that $$g(x;K,\lambda) = \left(1-r\left(1-\tanh^{2}\left(\frac{\lambda}{2}\right)\right)\left(1-\left(\cos^{2}\left(\frac{\pi x}{K}\right)\right)^{\frac{K^{2}}{4}}\right)\right)m\left(0;K,\lambda\right)$$ is an upper bound of $m(x;K,\lambda)$ when $r=0.89$.
Consider two cases: 1) $K=2$ and 2) $K>2$.
In the case of $K=2$, through use of trigonometric identities, it can be shown that $g$ is an exact expression for $m$ when $r=1$. By setting $r=0.89<1$, $g$ can only increase given that $r$ multiplies a positive-valued function. Therefore, setting $r=0.89$ leads to $g$ being an upper bound for $m$ in the case of $K=2$.
In the case of $K>2$, we will establish that $1-\left(\cos^{2}\left(\frac{\pi x}{K}\right)\right)^{\frac{K^{2}}{4}}$ is a \emph{lower bound} of
\begin{align}
\label{eq:lam-indep}
    \frac{1-g(x;K,\lambda)}{r(1-\tanh^2(\lambda/2))}.
\end{align}
First, we use the fact that the above function is monotonically decreasing in $\lambda>0$ and the fact that $\lambda$ is upper bounded, to set $\lambda$ to this value and establish a new $\lambda$-independent function of $K$ and $x$ that is a lower bound.
Since, as set in the main text, 
$K=\max\{10\lfloor\frac{1}{10(2\lambda+3/2J)}\rfloor,2\}\geq \frac{1}{2\lambda+3/2J}\geq \frac{1}{2\lambda}$, we have $\lambda \geq 1/2K$.
Since Eq. \ref{eq:lam-indep} is monotonically decreasing in $\lambda$, setting $\lambda$ to its maximum value of $1/2K$ yields a lower bound for Eq. \ref{eq:lam-indep} (on the valid domain of $\lambda$) that is only a function of $K$ and $x$.
It can then be numerically established that for all integer $K\geq 2$, we have
\begin{align}
   \frac{1-g(x;K,\lambda)}{r(1-\tanh^2(\lambda/2))}\geq\frac{1-g(x;K,1/2K)}{r(1-\tanh^2(1/4K))}\geq 1-\cos^{K^2/2}\left(\pi x/K\right)
\end{align}
This gives us the desired functional upper bound on $|f_j|^2$ as
\begin{align}
\label{eq:boundC}
    |f_j|^2\leq \frac{e^{-(K-1)\lambda}}{K^2}\left(\frac{\cosh(K\lambda)-1}{\cosh(\lambda)-1}\right)\left(1-0.89\left(1-\tanh^{2}\left(\frac{\lambda}{2}\right)\right)\left(1-\cos^{K^2/2}\left((\theta-2\pi j/J)/2\right)\right)\right)\equiv S(K,J,\lambda;j),
\end{align}

and this function $S(K,J,\lambda;j)$ decreases monotonically away from $j=J\theta/2\pi$ on the interval $-J/2\leq j-J\theta/2\pi\leq J/2$, which will be used later.

Putting the previous bounds together, we have that
\begin{align}
\frac{\sqrt{\textup{Var}(\hat{d})}}{\eta}&\leq 4\pi\frac{\sqrt{\mathbb{E}(|\hat{d}|^2)}}{{|f_{j*}|^2-|f_j|^2}}\leq 4\pi \frac{\sqrt{Q(K,J;j)}}{R(K,J,\lambda)-S(K,J,\lambda;j)},
\end{align}
where $Q$, $R$, and $S$ are defined in Eqs.~\eqref{eq:boundA}, \eqref{eq:boundB}, and \eqref{eq:boundC}, respectively.
This overall function is monotonically decreasing in 
$|j-J\theta/2\pi|$, and thus setting this value to its smallest allowed value of $|j-J\theta/2\pi|=1$ (the non-adjacent condition) leads to an upper bound that is independent of $j_*$ and $j$, 
\begin{align}
\frac{\textup{Var}(\hat{d})}{\eta^2}&\leq  16\pi^2 \frac{Q(K,J;J\theta/2\pi+1)}{(R(K,J,\lambda)-S(K,J,\lambda;J\theta/2\pi+1))^2}\equiv W(K,J,\lambda).
\end{align}
From this we are able to establish the final bound on the sufficient number of measurements to ensure an accurate estimate with high probability,
\begin{align}
\label{eq:explicit_M_bound}
    M\geq  128\pi^2\log(8J/\delta)\frac{Q(K,J;J\theta/2\pi+1)}{(R(K,J,\lambda)-S(K,J,\lambda;J\theta/2\pi+1))^2},
\end{align}
where $Q$, $R$, and $S$ are defined in Eqs.~\eqref{eq:boundA}, \eqref{eq:boundB}, and \eqref{eq:boundC}, respectively.
We note that, although $\theta$ appears in the arguments of $Q$ and $S$, this cancels with the $\theta$ in the definition, making the overall expression independent of $\theta$.

\section{Comparison to cost of standard quantum phase estimation}
\label{app:qpe_ft_deriavtion}

In this section, we analyze the cost of the standard quantum phase estimation (QPE) algorithm as described in \cite{cleve1998quantum}. This allows us to compare the fault-tolerant overhead associated with the traditional approach to the frugal approach that we have taken using the RFE algorithm.

Based on Ref.~\cite{cleve1998quantum}, the standard QPE algorithm can achieve an $\epsilon$-accurate estimation with probability greater than $1-\delta'$ by using $n = \lceil \log_2(\frac{1}{\epsilon})\rceil+\lceil \log_2(\frac{1}{2\delta}+\frac{1}{2})\rceil$ ancillary qubits and performing $2^{n+1}-1$ $c$-$U$ operations in the circuit. We note that here we have ignored the cost of the quantum Fourier transform operations, as in most cases it is negligible compared to the rest of the circuit. 
While this analysis assumes that the circuit is implemented perfectly, in practice, there will be some implementation failure probability.

The standard approach for analyzing the failure rate of quantum algorithms is to upper bound the failure rate as 
\begin{align}
    \delta \leq \delta_{\textup{alg}}+\delta_{\textup{imp}}.
\end{align}
Here, the implementation failure probability $\delta_{\textup{imp}}$ accounts for any uncorrected errors within the fault-tolerant protocols, contributing to the overall algorithm failure probability $\delta$ alongside the standard algorithm failure probability $\delta_{\textup{alg}}$.
This assumption holds reasonably well in terms of asymptotic scaling, with the fault-tolerant overhead exhibiting a logarithmic dependency on the inverse failure rate.
However, in the context of early fault-tolerant systems where resources are scarce, a more economical allocation of resources becomes advantageous.

Given a target total failure rate of $\delta$, the allocation of the error budget between the algorithm failure and the implementation failure becomes a crucial decision.
Though not optimal, a nearly optimal approach is to evenly distribute an error budget of $\delta/2$ to both failure modes.
Consequently, this establishes a lower bound on the required number of ancilla qubits,
\begin{align}
    n\geq \lceil \log_2(\frac{1}{\epsilon})\rceil+\lceil \log_2(\frac{1}{\delta}+\frac{1}{2})\rceil.
\end{align}
Similarly, the number of $c$-$U$ operations in the circuit satisfies the lower bound
\begin{align}
    \textup{\#}(c-U) \geq \frac{1}{\epsilon}\left(\frac{1}{\delta}+\frac{1}{2}\right)-1.
\end{align}
We will only slightly over-count resources by dropping the $-1$ above.
In order to ensure that $\delta_{\textup{imp}}\leq\delta/2$, each $c$-$U$ component must fail with a probability no greater than
\begin{align}
    \delta_{c-U}\leq \frac{\delta}{2}\frac{1}{2^{n+1}-1}\leq \frac{\delta}{2}2^{-(n+1)} = \frac{\epsilon \delta^2}{1+\delta}.
    \label{eq:cU_max_fail_prob}
\end{align}

We note that here the implementation error solely arises from QEC failure. As a reminder, the logical error rate for a surface code is assumed by the model $p_{\textup{logical}}=Ae^{-Bd}$ (as discussed in \S\ref{subsec:ft_model}), where $A$ and $B$ are constants from empirical observation of the surface code scaling and $d$ is the distance of the surface code. Assuming that each logical $c$-$U$ operation includes $\sim ND$ physical operations, where $N$ is the number of qubits and $D$ the circuit depth, the failure probability of one logical c-U is then 
\begin{align}
    \delta_{c-U}\leq NDp_{\textup{logical}}.
    \label{eq:cU_logical_failure_rate}
\end{align}
To ensure that the $c$-$U$ logical failure rate in Eq.~\eqref{eq:cU_logical_failure_rate} is bounded by the maximal acceptable failure probability as required by the algorithm from Eq.~\eqref{eq:cU_max_fail_prob}, we can therefore bound the logical error rate $p_{\textup{logical}}$
\begin{align}
    p_{\textup{logical}}&\leq \frac{\epsilon \delta^2}{ND(1+\delta)},
\end{align}
whereby we can then bound the minimal distance require to reach the desired success probability $\delta$ for the traditional QPE algorithm
\begin{align}
    d\geq \frac{1}{B}\ln\left(\frac{AND(1+\delta)}{\epsilon\delta^2}\right).
\end{align}

In the case of a high physical error rate with $A=0.5$ and $B=1.6$, achieving a success probability of $\delta=10^{-2}$ and precision of $\epsilon=10^{-3}$ for a QPE problem instance of $N=100$ and $D=1000$ requires a minimal distance of $18$ with the traditional algorithm. This corresponds to approximately 64,800 physical qubits, currently surpasses the capabilities of both NISQ and EFTQC devices. 

To illustrate this, in Fig.~\ref{fig:ft_overhead}, we compare the runtime, measured in the number of error correction cycles, between QPE and RFE as a function of the required physical qubits $2Nd^2$. It is important to note that while RFE has a higher runtime upper bound, it offers a tradeoff between runtime and the required distance/number of physical qubits in the EFTQC regime (as indicated by the shaded green region). This tradeoff becomes particularly valuable for upcoming devices that can accommodate a few thousand qubits and support error-correcting codes of moderate distances.

In practice, the actual runtime of RFE is expected to be significantly lower than the calculated upper bound. Additionally, we note that the runtime cost can be effectively reduced through parallelization across multiple devices of smaller sizes, leveraging the statistical nature of the RFE algorithm.